# Engineering Blockchain Based Software Systems: Foundations, Survey, and Future Directions

**Mahdi Fahmideh**, University of Southern Queensland, Australia
**John Grundy**, Monash University, Australia
**Aakash Ahmad**, School of Computing and Communications, Lancaster University, Leipzig, Germany
**Jun Shen**, **Jun Yan**, University of Wollongong, Australia
**Davoud Mougouei**, University of Southern Queensland, Australia
**Peng Wang**, **Aditya Ghose**, **Anuradha Gunawardana**, University of Wollongong, Australia
**Uwe Aickelin**, University of Melbourne, Australia
**Babak Abedin**, Macquarie University, Australia

Many scientific and practical areas have shown increasing interest in reaping the benefits of blockchain technology to empower software systems. However, the unique characteristics and requirements associated with Blockchain Based Software (BBS) systems raise new challenges across the development lifecycle that entail an extensive improvement of conventional software engineering. This article presents a systematic literature review of the state-of-the-art in BBS engineering research from the perspective of the software engineering discipline. We characterize BBS engineering based on the key aspects of *theoretical foundations*, *processes*, *models*, and *roles*. Based on these aspects, we present a rich repertoire of development tasks, design principles, models, roles, challenges, and resolution techniques. The focus and depth of this survey not only give software engineering practitioners and researchers a consolidated body of knowledge about current BBS development but also underpin a starting point for further research in this field.

## 1 INTRODUCTION

Blockchain-based software uses concepts and technologies popularized by cryptocurrencies such as Bitcoin – highly decentralized, open transaction ledgers with immutable content. Further advances, such as smart contracts and faster transaction processing, have provided advantages including transparent operation, user anonymity, auditability, and high scalability. A global survey of more than 1400 senior executives and practitioners across different countries by Deloitte [1] reveals that many IT-based organizations have found compelling blockchain use cases in areas like finance, transport, healthcare, and manufacturing. A study by Chakraborty et al., [2] reports that over 3,000 blockchain software projects had been hosted on GitHub in March 2018 and that this number had been sharply doubled to nearly 6,800 in October of the same year.

However, in parallel with the growing interest in adopting Blockchain-Based Software (BBS) systems, the accounts of project failure and system attacks in this domain have also become noticeable. For example, the MtGox attack in 2014 led to a declared loss of $600 million, Bitfinex attack in 2016 led to a loss of $65 million, and a DAO (decentralized autonomous organization) attack in 2016 caused the withdrawal of Ether digital currency funds worth $50-60 million. These failures are commonly attributed to a lack of adoption of a *systematic development approach*, a project management flaw that is aptly phrased as *un-ruled and hurried software development* by Porru and Pinna [3]. As such, blockchain projects in some organizations are in the experimental mode. A Gartner report in 2019 predicted that some blockchain project initiatives will be discontinued in the near future due to these development failures [4]. These negative outcomes of blockchain applications have been reflected in another Gartner report [5] which mentions that *many chief information officers overestimate the capabilities and short-term benefits of blockchain […] thus creating unrealistic expectations from blockchain platform vendors and service providers [….]*. Given these shreds of evidence, there is skepticism about whether blockchain technology really fortifies software systems. Research agendas, development frameworks, and prioritization of futuristic work have been put forward [6],[7].

To alleviate these issues, there is a need for more systematic software engineering approaches, sometimes deprecatingly referred to as engineering methodologies, to ensure the quality of both development and maintenance of BBS systems, which has been corroborated by prior research [3]. In this spirit, Chakraborty et al. [8] surveyed 156 active blockchain software developers in 145 blockchain projects hosted on GitHub and concluded that conventional software engineering methods and best practices need to be adapted to address the unique characteristics of BBS. These characteristics include a higher emphasis on security and reliability, promoting transparency in data and operations, consensus on the state of data, the immutability of system transactions, peer-to-peer distributed network, and the need for in-depth knowledge of cryptography, to name a few [3],[9]. Some researchers envision blockchain *as another class of software like the Internet* [10] and call for revisiting software engineering attuned with the characteristics of this technology [2],[3],[11].

In recent decades, the software engineering discipline has evolved to meet industry needs and to keep abreast of modern computing technologies that have influenced the development and adoption of software systems. In line with this, there is a growing awareness that the end-to-end development lifecycle of BBS requires new innovative guiding engineering approaches to bind together technical programming models, platforms (e.g., Ethereum), and technologies (e.g., Bitcoin scripting languages) [2],[3],[12]. Although small BBS projects may be manageable in an ad-hoc manner, a systematic engineering approach becomes essential if a BBS project is large and the target BBS is aimed to support the core business processes of an organization. BBS components deployed on blockchain platforms are targeted to be attacked by malicious entities. A systematic development approach can guide important activities, guidelines, checklists, design principles, and heuristics to mitigate, trace, and rectify the cause of vulnerabilities and errors. Hence, there is a need for a literature survey answering the looming questions like *how software engineering for the development of BBS systems is perceived according to the current state-of-the-art and industry efforts*?

Despite the growth and interest in the blockchain field, the aspect of software engineering associated with developing this class of systems remains under-explored and under-conceptualized [2],[3]. Numerous studies are available in the literature including proposals, methods, empirical studies, and experience reports discussing different questions related to BBS development [13], [14]. The classification and analysis of these different studies is a challenge for researchers and practitioners, which demands conducting a rigorous and well-defined literature review. For this reason, we employed the search and selection protocols proposed by the procedure of *Systematic Literature Review* (SLR) [15] and qualitative data analysis based on *Grounded Theory* [16] to provide a complete review of the state-of-the-art of BBS development. Our literature review considers 58 selected core studies [17] and they are organized into a new conceptual framework. The framework addresses different aspects of BBS development, namely, *approaches*, *processes*, *modeling*, and *role* which guide software developers, business managers, and academic researchers in the exploration of practical side, implications, and challenges related to BBS development. To the best of our knowledge, our survey is an unprecedented work and a novel attempt at summarizing the state-of-the-art research and practice on the end-to-end engineering lifecycle for BBS. As discussed later, no other published surveys match the scope, depth, and coverage of our survey and its findings. The key contributions of this work are:

— a coherent set of method fragments grounded in the literature to facilitate a better understanding of the BBS engineering lifecycle;
— a guidance framework, as *having a tail-light to follow*, for novice software practitioners engaging in BBS development context; and
— an outlook on open challenges in BBS development for future research directions.

The rest of this paper is organized as follows. In Section 2, the discussion of blockchain concepts lays the basis to raise essential implications for the development of BBS. Section 3 summarizes past literature work related to our current survey. In Section 4, we introduce our proposed conceptual framework, which is derived from the extant literature on BBS engineering and provides a lens for an analytical literature review. Section 5 delineates our findings based on our framework, followed by a discussion on existing knowledge gaps in the literature in Section 6. We close with a discussion of our survey limitations in Section 7 and a concluding remark in Section 8.

## 2 BACKGROUND

This section introduces the core concepts including the building blocks for a typical BBS (Section 2.1) and the necessity of software engineering for BBS (Section 2.2). The concepts and terminologies introduced in this section are used throughout this paper to guide the discussion of the survey results.

### 2.1 Building Blocks for Blockchain

New system engineering approaches are often grounded on fundamental concepts and underlying logic. As depicted in Figure 1, blockchain is commonly rooted in a few core elements that a BBS is built upon [10], [12], [18], [19]. A *distributed ledger* is a form of a shared database that may exist across multiple locations and among several participants [12]. It enables parties to authenticate, process, and validate transactions without the need for central authority or intermediary. *Blockchain*, with its origin in the Bitcoin cryptocurrency [18], is a type of distributed ledger. It is viewed as a promising initiative for the secured and reliable next generation of Internet-Based information technologies [19]. Some industrial domains have already developed BBS and others are still figuring out a reasonable use case to offer new BBS for transactional digital services [12]. In a simple view, blockchain is an accounting book or digital distributed database. It has a chain of *blocks* (i.e., records) that are sequentially linked together. Each block depends on its predecessor block and is secured via cryptography techniques [19]. A block contains transactional data, a time stamp, and a hash value of its previous block. Also, a block saves an arbitrary set of transactional data that is created by a *node* (a computer in the distributed ledger network). The chain of blocks is stored on a distributed network of nodes where each node contains a copy of the entire blockchain. The chain is visible and verifiable by all nodes participating in the network. Once a block with its own time-stamp is appended to the chain, the creator node broadcasts that block to all the other nodes in the peer-to-peer distributed network. Once nodes receive the block, they validate it via predefined check and add the block to their own local blockchain copy to provide a single source of truth. As the name implies, some nodes, called *validators*, are responsible for validating a newly added block to the blockchain. The data records in a block are non-reversible, transparent, and become an immutable part of blockchain after they are accepted by all nodes. Such a chain of blocks provides a secured means of information exchange between systems without involving a trusted third party and is suitable for record-keeping operations such as financial transactions, medical records, and so on [12]. Casino et al. [20] predict that blockchain technology will continue to be adopted by multiple industries such as manufacturing, cybersecurity, 5G networks, and IoT to augment security, track records, and ensure data consistency.

A key element of blockchain technology is the ability to create *smart contracts*. The term, which was coined by Nick Szabo in the mid-1990s [21], is defined as translating the clauses of a business contract into code and embedding them into software or hardware to make them automated and self-execute. This reduces the cost of contracting between transacting parties and avoids malicious actions during contract execution. Different alternative definitions of smart contracts have been already proposed. For example, Crosby and Pattanayak [18], view a smart contract as database slots that reserve the necessary logic for the creation and validation of transactions and enable users to read, update, and delete the data that is stored in blockchain platforms. A broad definition of smart contract that covers the breadth of variant definitions is proposed by Clack et al. [22]. That is, a smart contract is *an automatable and enforceable agreement. Automatable by computer, although some parts may require human input and control. Enforceable either by legal enforcement of rights and obligations or via tamper-proof execution of computer code* [22]. Smart contracts are either implemented via domain-specific programming languages, e.g., Solidity on Ethereum platform, Liquidity on the Tezos platform or by general-purpose programming languages, e.g., Java, Go, Kotlin to ease smart contract implementation. For instance, Ethereum platform supports the built-in language Solidity and executes the compiled bytecode of Solidity scripts on its execution environment, called Ethereum Virtual Machine (EVM). A smart contract's script code is invoked if it receives a transaction from a user. The smart contract defines necessary entry points and triggers of transaction execution. In the terminology of Ethereum, each entry point is called a *function*. A transaction specifies entry points at which function codes of a smart contract should be executed. In line with this, a BBS is developed by codifying business logic into smart contracts. The smart contracts are then deployed and autonomously executed on decentralized ledgers, e.g., Ethereum. This end-to-end development needs a systematic development lifecycle or software engineering methodology [3] as discussed next.

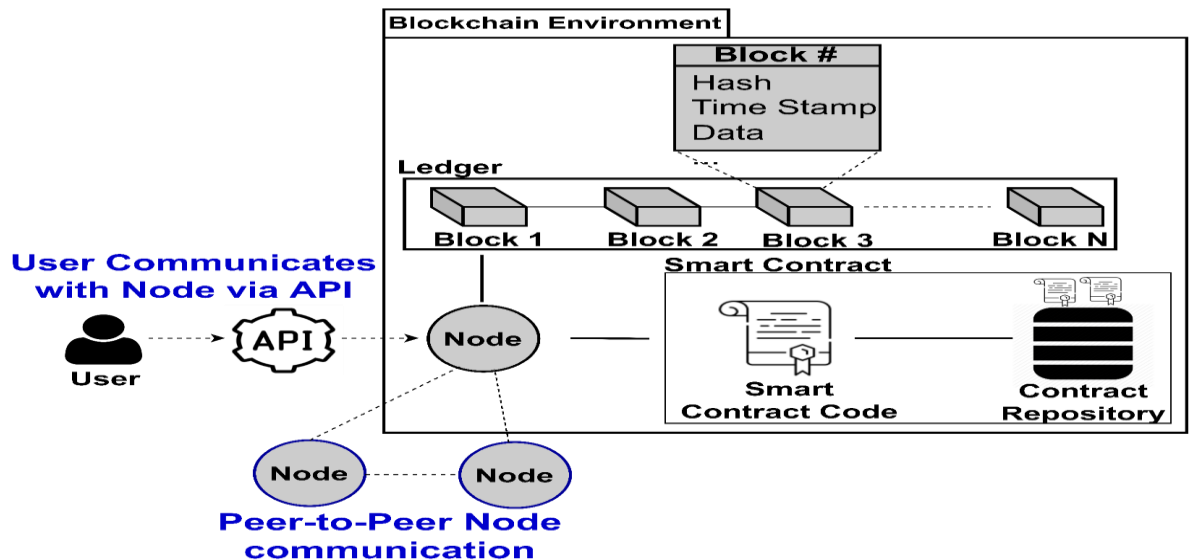

Fig. 1. Fundamental constituents of the blockchain architecture

## 2.2 Software Engineering for BBS system

We first establish an understanding of software engineering for BBS systems. A simple definition, according to Porru et al., [3], is that BBS *is a new type of software that uses the implementation of blockchain in its components*. If a system depends on the blockchain, its development may *revolve around this technology* [3] and be perceived as *actually different from a non-BBS development* [2]. Unlike non-BBS systems, the development of BBS systems raises new challenges that software engineers have to cope with. As explained in Section 5, these challenges highlight the importance of further research in this field in both technical and non-technical elements, such as better guidance for blockchain developers, a trade-off between security and performance, the choice of consensus protocols in relation to transaction time and computing power requirements, the cooperation of multiple institutions and stakeholders, [2],[7],[19]. Some of these challenges can be attributed to the immaturity of the technology itself as it is common in all new technologies at their inception. Others are intrinsic to blockchain concepts and characteristics. Conventional wisdom in software engineering acclaims that the appropriate use of engineering methodologies ensures the development of quality software that meets expected outcomes while satisfying budget and timeframe constraints [23]. Inspired by the lessons learned in the software engineering industry informing that *the more a system gets complex, the more it is open to total breakdown* [24], it can be suggested that understanding and adopting systematic engineering approaches for BBS reduces the likelihood of user dissatisfaction and unforeseen security and other errors that may become costly to fix. Drawing on the common prudence in the software engineering field [25],[26], without a systematic approach it will be hard for a software team to dissect and trace the cause of errors and correct them when things go wrong. In other words, the quality of development significantly impacts the quality of the resulting BBS running on blockchain platforms.

## 3 RELATED SURVEYS

Many literature reviews on different aspects of the blockchain topic are available. As shown in Table 1, we divided the related studies into *widely related surveys* focusing on software engineering for Internet-based distributed computing applications and *closely related surveys* pertinent to blockchain applications.

## 3.1 Surveys on Engineering Internet-Based Computing Systems

Widely related surveys focus on intersections and synergies between Internet-based computing technologies such as blockchain, service computing, cloud computing, and IoTs. They are theoretically grounded on the distributed architecture that provides a backbone and virtually unlimited computational resources and on-demand services to develop and deploy software systems. Surveys of Fahmideh et al., [27], [28], Lane et al., [29], and Razavian and Lago [30] are categorized under the widely related surveys. For example, cloud computing empowers the scalability and performance of blockchain platforms by providing service delivery models such as Infrastructure as a Service (IaaS), Software as a Service (SaaS), or Platform as a Service (PaaS). Fahmideh et al. in their initial work [31] that is later elaborated in [27] compare the characteristics of 43 reengineering approaches, published between 2007 and 2015, to make legacy software systems cloud-enabled. Their survey synopsizes the status quo and presents a rich set of development tasks and recommendations commonly incorporated into typical migration processes to cloud platforms upon service delivery models such as IaaS, SaaS, or PaaS. Blockchain is used as an enabling technology for IoT-based applications as it provides a secured decentralized network for data management and communication of sensors and devices. Fahmideh and Zowghi [28] present a generic development process lifecycle for IoT-based applications by mapping 63 approaches, published between 2008 and 2019. They conclude that development roles, requirements analysis, modelling, testing, and tailorability of development processes are not addressed in the existing approaches to engineer IoT-based applications.

Razavian and Lago [30] present an analysis of reengineering processes in 75 approaches for developing SOA (service-oriented architecture) based software applications. Their main contribution is to devise a new conceptual model and classifications of activities such as code analysis, architecture recovery, service design, and implementation in order to integrate with and componentize legacy applications to web services. They identify major challenges of (i) value creation for service-based software engineering, (ii) decision making on tool selection, and (iii) legacy understanding without reverse-engineering.

In addition, a possible interplay between blockchain and big data is to use blockchain for storing big data analytics software applications that enable user authentication, recording data access history, and proper use of encrypted data on peer-to-peer distributed networks. Davoudian and Liu [32] review challenges related to three major software engineering activities of requirements analysis, design, and implementation in the context of big data analytics development based on academic and multi-vocal literature. Despite the focus on software engineering of systems relying on Internet-based computing technologies, all the above surveys do not directly deal with technical implications and complexity of software engineering aspects highlighted in BBS such as adopted approaches, processes, models, and roles that represent the multifaceted view of the development lifecycle as shown in our framework. Despite their usefulness in forming a basis for our conceptual framework, we deem that the recent literature surveys on conventional software development processes and improvement, e.g., [33], do not overlap with the focus and scope of our research objectives.

### 3.2 Surveys on Blockchain-based Systems Engineering

Scant work is available as being exclusively devoted to a review of existing proposals on the software engineering for BBS. To support in-depth technical analysis, we discard introductory surveys that aimed at demystifying the notion of the blockchain or discussing prevailing challenges in adopting this technology in software applications and organizations. The reason is that they do not concentrate on the aspect of the development lifecycle. Despite their usefulness in crafting our research objective, introductory surveys fall outside the scope and focus of this survey. Some example surveys in this genre are the taxonomy-based surveys on the blockchain usage trend in the IoT context [34], consensus algorithms and application domains of blockchain [19], security concerns in blockchain adoption [35], [36], and business applications of blockchain [37]. The most extensive work is provided by Yli-Huumo et al., [13] with a research objective to *understand the current research topics, challenges, and future directions regarding blockchain technology from the technical perspective*, (page.1). By analyzing 41 primary studies, published between 2012 and 2015, where 80% of the papers focus on Bitcoin systems and less than 20% deal with other application domains, Yli-Huumo et al. highlight that the lack of concrete evaluation metrics and techniques to measure scalability efficiency in terms of throughput and latency is left unaddressed in the literature. The goal of the survey by Liu et al., [14], on the other hand, is to investigate the design of smart contracts and the implementation of trusted transactions among parties without mediators. They selected 53 papers to show the state-of-the-art of this topic, developed a taxonomy towards the security verification of blockchain smart contracts, and discussed the pros and cons of each category of the related studies. Vacca et al. [38] review 96 studies published between 2016 and 2020 to identify existing solutions in tackling software engineering-specific challenges related to the development, test, and security assessment of blockchain-oriented software.

None of the surveys in Table 1 is specifically geared towards the development lifecycle for BBS and intrinsic characteristics of blockchain that challenge well-established development processes of conventional software systems. As a response, we have narrowed our focus to extant studies that propose completely or partially, end-to-end approaches for BBS development. This includes development themes, essential tasks for incorporation into the development process, emerging roles, and modeling challenges. From this angle, our survey is more precise when compared to the related surveys. Our literature analysis is performed through a conceptual framework, encompassing four important dimensions of development approaches, processes, modeling, and roles that are not covered in related surveys. Our study aims to aid software teams, managers, and researchers in characterizing the extant material in the literature and mastering skills necessary to develop BBS. Due to the different focus and scope of our review, many research papers that we discuss in this survey are not covered by these existing surveys. For instance, none of the studies reviewed in our survey is covered by [14]. Our work supersedes the existing surveys as it considers many recently published studies in blockchain literature that are not included in the related surveys.

Table 1. Comparison of proposed survey with existing related surveys

| | Survey reference | Computing paradigm | Research focus | Publication year | Reviewed studies | Study type |
|---|---|---|---|---|---|---|
| Widely related surveys | Razavian and Lago [30] | SOA | Reengineering processes of legacy software systems to SOA | 2015 | 75 | Systematic literature review |
| | Fahmideh et al. [9] | Cloud | Reengineering processes of legacy software systems to cloud computing platforms | 2016 | 43 | Systematic literature review |
| | Davoudian and Liu [32] | Big data analytics | Challenges associated with software engineering activities of requirements analysis, design and implementation of big data analytics software applications | 2020 | Not stated | Literature survey |
| | Fahmideh and Zowghi [28] | IoT | Development processes for IoT based systems and platforms | 2019 | 63 | Systematic literature review |
| Closely related surveys | Panarello et al. [34] | IoT and blockchain | Integration of blockchain and IoT applications in terms of different application domains, usage patterns, device manipulation, and data management | 2018 | Not stated | Literature survey |
| | Zheng et al. [19] | Blockchain | Blockchain architecture, key characteristics of the blockchain, consensus algorithms and protocols | 2018 | Not stated | Literature survey |
| | Mohanta et al. [36] | Blockchain | A comprehensive analysis on blockchain applications including implementation challenges associated security and privacy | 2019 | 135 | Literature survey |
| | Li et al. [35] | Blockchain | Real security attacks to popular blockchain systems | 2020 | Not stated | Literature survey |
| | Konstantinidis et.al [37] | Blockchain | Business applications of blockchain in both public and private sectors | 2018 | 44 | Systematic literature review |
| | Liu et al. [14] | Blockchain | Security assurance and correctness verification of smart contracts | 2019 | 54 | Literature survey |
| | Vacca et al. [38] | Blockchain | Challenges in implementing, testing, and secure-aware smart contracts | 2021 | 96 | Systematic literature review |
| | Our survey | Blockchain | Blockchain software engineering | - | 59 | Systematic literature review |

## 4 RESEARCH APPROACH

Due to reasons of (i) the sheer volume of published material in an overgrown and unkempt domain like blockchain and (ii) the exploratory nature of this literature review, we employed kernel techniques from software engineering and information system research fields. In particular, we used the guidelines for conducting a *Systematic Literature Review (SLR)* [15] and qualitative data analysis based on *Grounded Theory* [16] as outlined in Figure 2 and detailed in sections 4.1 and 4.2. We want to provide a conceptual framework, encompassing commonly occurring fragments of software engineering for BBS and their interrelations. In the following subsections, we describe our literature review procedure and the steps to derive the conceptual framework, guided by the illustration in Figure 2.

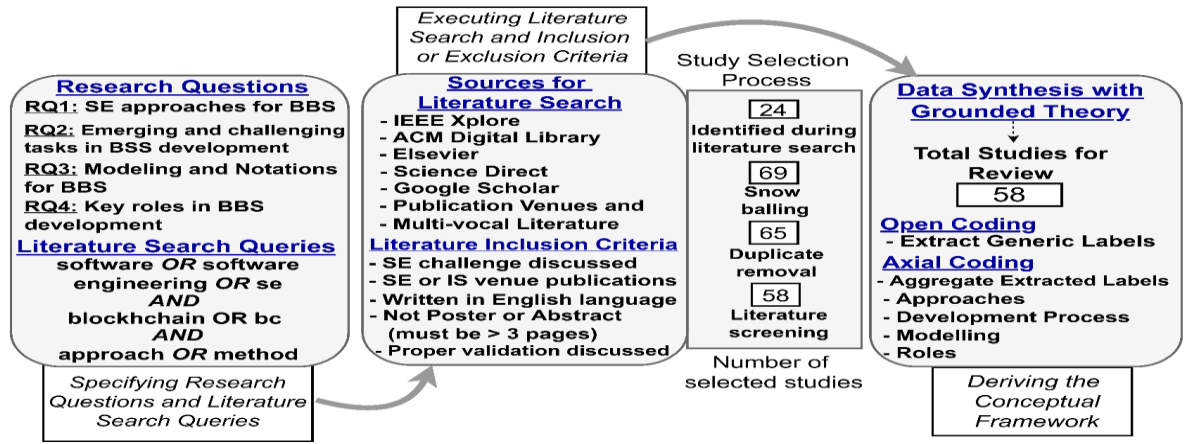

Fig. 2. Sequence of steps undertaken in our literature survey

### 4.1 Systematic Literature Review

As stated earlier, this study aims to portray the current research landscape in software engineering for BBS. We formulated the following research questions:

- *RQ1: what are the common software engineering approaches that are adopted for BBS development?*
- *RQ2: what are the key emerging and challenging tasks in developing BBS systems?*
- *RQ3: what software modeling approaches and notations are applicable in BBS development lifecycle?*
- *RQ4: what are key roles in a BBS development endeavor and how do they play?*

We followed the well-accepted procedure and guidelines by Kitchenham et. al. [15] to conduct this SLR as they deem to be an appropriate way to systematically identify and qualitatively specify the criteria for the selected studies involved in the review and data synthesis. In the blockchain field, the concepts, nomenclature, and terminologies have not been consolidated yet. Hence, our SLR undertaking was not a linear and mechanical procedure. Instead, our review tended to be hermeneutic and iterative. In other words, to get immersed in the literature, as recommended by Brun et al. [39], we read introductory and literature survey papers, some of which are listed in Section 3 (Table 1), to get an in-depth understanding of the important aspects of BBS engineering and further refinements of the literature review, in particular, inclusion/exclusion criteria definition and data extraction step. Our SLR's key steps included review protocol, search queries, selection of study sources, study selection, data extraction, and data analysis as shown in Figure 2. More exactly, we determined the main terms *software engineering*, *blockchain*, and *approach* based on the research questions RQ1-RQ4 to define a set of search queries along with related abbreviations, plurals, and extended them with alternative synonyms. The *approach* was also used because some papers from our initial literature search had used these terms to describe the frameworks and methods for the development of BBS. Search queries were executed against the main scientific digital libraries including IEEE Explore, ACM Digital Library, Elsevier, SpringerLink, and ScienceDirect. We then selected papers that met the following inclusion criteria:

- explicitly discussed software engineering challenges related to BBS;
- published in the international software engineering and information systems venues such as journals, conferences, congresses, and workshops as well as computing specific venues, e.g., computer networks and sensor-driven systems, which streamline software aspects of BBS;
- was written in the English language with at least three pages (not an extended abstract), and
- provided a proper validation of their work such as case study application, exemplar scenario, interview/survey of practitioners of blockchain practitioners, simulation, comparative analysis, or theoretical evaluation.

Thus, based on the criterion above and SLR practice we discarded studies that were

- opinions, white papers or short papers without any sorts of evaluation results;
- providing only background on the blockchain or discussed the applications of blockchain in other enabling technologies like IoT, big data, and cloud computing;
- Thesis or other non-refereed publications; and
- Not written in the English language

In addition to the mentioned digital libraries, we paid attention to peer-reviewed venues exclusively dedicated to blockchain publications themes such as *International Conference on Blockchain* and *International Conference on* *blockchain and Trustworthy Systems*. We also sought online non-academic literature, i.e., multi-vocal literature, such as internet blogs and trade journal articles, which could share real-world findings surrounding BBS development. A total of 24 papers were identified from the previous search iterations in the selected digital libraries. An important technique that we rigorously used during the literature search iterations was snowballing through which we identified 65 new important papers from the reference sections of the initially identified papers. In the last search iteration, snowballing technique resulted in 69 papers that were reduced to 65 after removing 4 duplicated papers from the same authors. We screened the title, abstract, and preferably full texts of the papers, and shortlisted 60 papers after applying the inclusion criteria. To avoid being biased, the list of the identified papers by the main research investigator of this survey was double-checked by one co-author, independently. Two papers were removed after they were found unsatisfactory to the inclusion criteria and the objective of this survey. The literature search iteration in its last round culminated in 58 papers listed in [17]. A summary of the identified papers' demographic information (e.g., types and years of publications, types of validations, and geo-distribution of authors) is shown in[17]. We denoted an identity to distinctly refer to each selected paper, which is the appending the letter S and a number (i.e., [Sn], n is unique a number from 1-58). These identities are used throughout this survey and conceptual graphs in Section 5.

**4.2   Conceptual Framework for Software Engineering of BBS**

In light of different terminologies and broad varieties of blockchain concepts, it is essential to first establish a foundation to enable our literature analysis pertinent to software engineering for BBS. To this end, we devised a conceptual framework that, according to Miles and Huberman [40], would enable researchers to graphically explain the main concepts and relationships in a domain of interest. In this survey, the derivation of such a framework was based on Grounded Theory [16], which has been found useful in qualitative research and its application has been well-recognized in several fields including medical sociology, education, and management. In recent years, using Grounded Theory in the software engineering field has received growing attention [41]. It aims to generate theories and conceptual models out of a collection of data about a research problem and to enable researchers to have amenable interpretation and free comprehension of data in different ways. The data can be sourced from, for instance, field observations, interviews, secondary data, semi-structured data, pictures, and diagrams. The generation of the theory from the data is based on a few steps including *open coding*, *axial/selective coding*, and *theoretical coding* that can be, according to Strauss et al. [42], governed and consulted by the pre-defined concepts and views from domain literature to enhance the theoretical sensitivity. We took into account the literature on conventional software engineering [43] to analyze the identified papers. We used open coding to break down and label data into smaller fragments [42] to analyze and highlight the text segments in each selected paper that could be considered as a discrete piece of data annotated with a descriptive *label*, as a series of high-level *intellectual bins* [40]. The labels could help structure and focus on the context of the data in the papers. As an example, to demonstrate the inner working of open coding, Table 2 presents an excerpt of text segments that were extracted from the selected papers [S24], [S25], [S26] (see [17]). When these quotes in Table 2 are collectively viewed, their common theme, despite differences in wording, is related to the notion of the *suitability of blockchain*, which should be performed by software teams at the early stages of the development process. Thus, we defined a label in the conceptual graph of the framework as a development task and named it as *feasibility analysis of BBS* supported by the identified papers. These will be delineated in Section 5.2.1.

Table 2. An example of emerging development task named *feasibility analysis of BBS* via open coding

| Quote from the source paper | Source |
|---|---|
| *This technology is being applied to an increasing range of industries and problem spaces. However, such an application might not always be appropriate or optimal; in many cases [...] while evaluating potential applications of blockchain at CableLabs, [...] I developed a framework for determining whether blockchain is appropriate [...]* | [S24] |
| *Investigating the feasibility of developing a smart contract that manages a network of banks [...] is a big challenge, above all looking at the throughput of the blockchain and hence to the rate at which a blockchain can confirm transactions.* | [S25] |
| *Due to the long commit time and high transaction fees on a public blockchain (where fees are largely independent of the transacted amount), it is often infeasible to store every micro-payment transaction on the blockchain network. On-chain transactions are suitable for transactions with medium to large monetary value, relative to the transaction fee.* | [S26] |

The open coding step resulted in labels such as *requirements analysis for BBS*, *state management design*, *replication and synchronization design*, *authentication and authorization design*, *interaction design*, *smart contract design*, *consensus mechanism design*, and *incentive mechanism design* each of which is supported by at least one reference to the selected studies. Once labels emerged over 58 identified papers as the source of data, axial coding was performed to aggregate and condense related labels to form broader groups of labels. We derived four essential groups of labels,

a.k.a. aspects, namely (i) *approaches*, i.e., software engineering foundations that could govern the end-to-end BBS development lifecycle, (ii) *development process*, (iii) *modelling*, and (v) *roles*. More specifically, some labels were grouped as a kind of *modelling* aspect: a *smart contract model*, recommended by studies [S1], [S5], [S31], [S46], [S56], and [S57], which is used to represent the expected functionalities and features of smart contracts to be implemented and tested by low-level code scripts running on blockchain platforms.

Axial coding has resulted in labels related to the aspect of the *role*, which are involved during a BBS development endeavor. If a label couldn't be classified under any or indirectly related to one or more of these aspects, we could define a new label or group it under a related aspect. In other words, some labels were related to decision factors for incorporation into the analysis and design activities, e.g., *decision on-chain and off-chain components*, *decision on blockchain type*, and concerns e.g., *visibility and transparency*, *organization restructuring*, *uncertain status*. Axial coding was further elaborated, in line with the literature on conventional software engineering [43], to create new high-level labels that could classify fine granular labels with common themes under the same aspect. For example, the tasks *feasibility analysis of BBS* and *requirements analysis for BBS* were subsumed under the *BBS analysis phase* in an analogy with the analysis phase in conventional software engineering. Often, there should be references supporting labels in this conceptual framework generated from Grounded Theory. These references to the identified papers are listed in the conceptual framework's graphs and cited in Section 5 wherever we present the results of the literature review. It should be noted that a paper could belong to multiple aspects of the framework. Grounded Theory was ended with theoretical coding where we established the conceptual relations between substantive labels, resulting in the conceptual framework for BBS engineering.

Using the Grounded Theory in this literature survey was not intended to develop testable hypotheses nor to answer unexplored research queries in the blockchain literature. Rather, we used it to (i) analyze and compare the extant papers in the literature on BBS engineering, (ii) identify existing engineering challenges and corresponding resolution techniques, and (iii) construct further research directions for unresolved challenges. Our generic conceptual framework, yet pluralistic and platform-independent, binds the essential aspects together, provides an integrated overarching view of software engineering for BBS to facilitate understanding of existing challenges and solutions proposed by the blockchain community. It also enables further research and development in this field. The next section presents the framework that was produced to structure a coherent set of relevant aspects of BBS engineering.

## 5   RESULTS

To present our results in an illustrative manner, we use visual conceptual framework diagrams as represented in Figures 3,4,7,8, and 9. In this conceptual framework, a *node*, either sub-node or parent node, may refer to a task/sub-task, technique, decision factor, challenge, modeling, or role that comes into play during a BBS development endeavor. *Directed edges* (or a*rrows)* show relationships among nodes. The nodes and arrows are supported by one or more source studies where the full demographic information of the study set is listed in [17]. As mentioned in Section 4, rooted in the fundamental aspects of software engineering, the diagrams in the framework characterize BBS development based on the four major aspects of the *approach*, *process*, *modeling*, and *roles*, which are, respectively, related to answering RQ1-RQ4 (Section 4.1). These allow software teams to explore and make a comparison between different BBS development options. These aspects and their importance are substantiated below.

The aspect of *development approach*, i.e., theoretical foundation, is rooted in the fact that software engineering endeavors are not impartial on their own, rather they have underlying development perspectives, including principles, assumptions, and drivers that govern and rationalize an end-to-end development lifecycle in a given project context [44]. BBS development is thus founded on some fundamental concepts and basic logic such as trust, decentralized and distributed public/private ledger, cryptography, smart contract, consensus algorithm, and transaction transparency that should be taken into account by software teams. These are critical and deserve close investigation.

The aspect of *process* specifies guidance for development lifecycle phases, necessary tasks, and techniques that are sequenced into the development of BBS. We borrowed the generic phases of the software development lifecycle (SDLC), i.e., *analysis*, *design*, *implementation and test*, and *maintenance*, introduced by Pressman [15] allowing us to organize and represent the software engineering process body of knowledge. The rationale to incept BBS development process like SDLC is that, regardless of an application domain or adopted underlying blockchain platforms, it is generally taken as axiomatic that engineering processes are almost the same at the macro and abstraction level, though they differ in fine-granular development tasks and technical-centric implementation details. This interpretation is consistent with previous exploratory literature surveys in other Internet-based computing domains such as IoT [28], [45], cloud migration processes [46], and SOA based development processes [47],[48] where the authors leverage SDLC phases to classify, compare, and contrast selected studies and to identify unaddressed research gaps. This aspect enables a better understanding of BBS life cycle management and it is populated with the identified commonly occurred development process tasks from the literature.

The aspect of *modelling* deals with the representation techniques and notations that software teams may employ for the outputs of development process tasks during a BBS development endeavor. This aspect, itself, encompasses two sub-aspects of (i) a *model*, a.k.a. *work-product/artefact*, as a result of performing development tasks according to Gonzalez-Perez et al. [49], and (ii) a *modelling language* to represent and maintain generated models. A model, as generally defined by ISO/IEC 24744 standard meta-model for software engineering methodologies [50], is an artefact of interest to a software team. In a blockchain project context, this may be, for example, a legal document, distributed ledger architecture, Bitcoin transactions, smart-contract template, Solidity code, or a piece of information related to the project. On the other hand, a modelling language provides underlying elements to represent a model. For example, class, inheritance, aggregation, or composition in the object-oriented software development paradigm enables developers to express different structural and behavioral facets of a model at a different level of abstraction serving as overlapping aspects of system development.

The *role* aspect indicates associated responsibilities, interactions, and required skills in a BBS development endeavor. This aspect in our framework is required as the development of BBS relies on different technical expertise, business vision, and delivery skills, which imply defining a development role acquisition plan. Bosu et al. [S22] suggest that special skills, beyond those known ones in non-BBS development, are needed for people considering joining BBS projects. There are advantages in defining the aspect of roles participating in BBS development. That is, specifying BBS roles and their professional responsibilities enable the separation of concerns so that if changes occur to one role, there will be less influence or dependency between the roles. Moreover, this clarifies the responsibilities of each project stakeholder, enables better team management, and monitors activities per role. Another advantage of defining the role aspect in our framework is to inform software teams who are with limited experience in blockchain and are not quite sure about the roles required in BBS development.

### 5.1 RQ1: What are the common software engineering approaches that are adopted for BBS development?

In the identification of proposed approaches for BBS development in the literature, we did not set any pre-defined classification during the open coding step of Grounded Theory. Instead, we found approaches based on what appeared in the selected studies. We observed five distinct themes, as depicted in Figure 3, that BBS development may leverage, namely (i) *Agile based development* (e.g., [S1],[S4],[S16],[S23],[S27],[S28],[S51]), (ii) *model-driven development (MDD)* (e.g., [S5],[S16],[S28],[S57],[S29],[S30],[S34],[S52]), (iii)*architecture-based BBS development* (e.g., [S3],[S4],[S11],[S12],[S33]), (iv) *pattern-based BBS development* (e.g., [S35],[S36],[S37],[S38],[S39],[S40]), and (v) *ontology-based BBS development* (e.g., [S41],[S42],[S43]). The adoption of these approaches is not mutually exclusive, rather they can be used together and orthogonally used across the development lifecycle. For example, in [S16], authors have shown both the model-driven and the Agile-based approaches for BBS development.

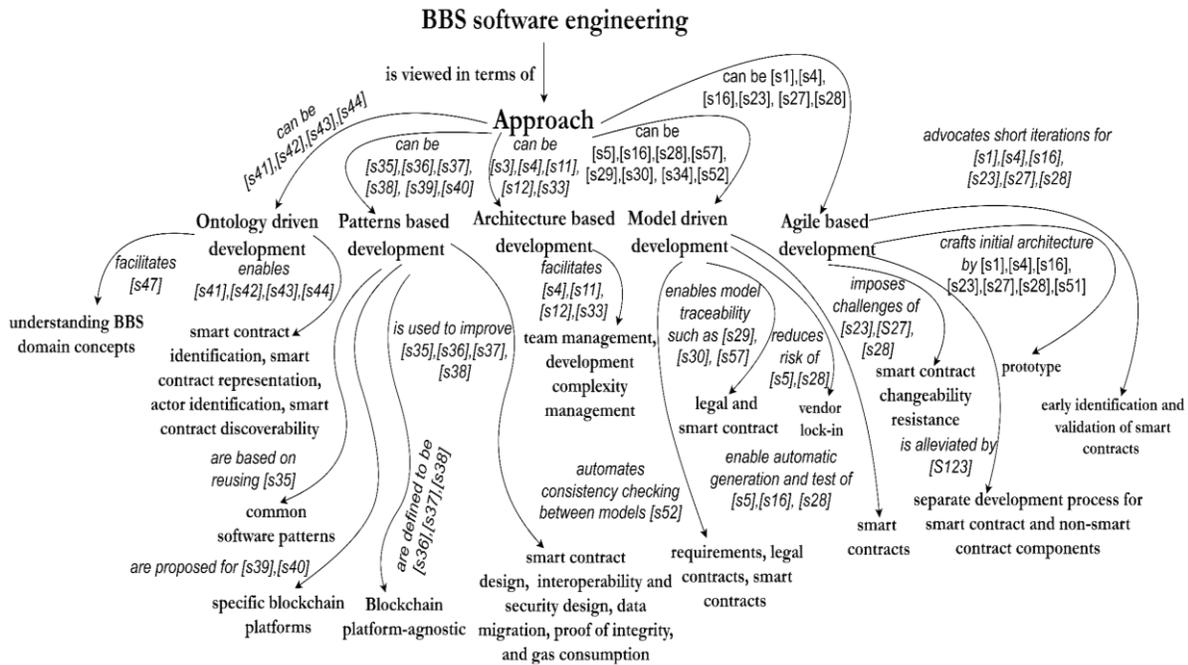

Fig. 3. *Approach* conceptual map

*5.1.1 Agile-based BBS development*

The common perception of agile-based software development is distilled in the *Agile Manifesto* [51] and used in well-known agile software development methodologies such as SCRUM, XP, and Lean Kanban. They emphasize practices such as active user involvement, lightweight modeling, short cycles, iterative releases with frequent customer review, and the priority of working software over documentation. Blockchain researchers have utilized agile practices for non-real-time critical BBS development [S1],[S4],[S16],[S23],[S27],[S28],[S51]. Agile prototyping, a.k.a. Spike Solution in XP, has been found to be a useful technique at the early stage of BBS development for the purpose of (i) requirements elicitation and specification from target blockchain platforms, (ii) forming a base target BBS architecture without dependency on a specific blockchain platform, and (iii) identifying uncertainties in system quality factors such as transaction execution performance, security, as well as the trade-off between them [S1],[S4],[S28]. Marchesi et al., [S23] discuss the usefulness of *user stories*, i.e., defining target BBS features as seen from stakeholders' point of view, to get early feedback from stakeholders to validate required smart contracts. Being open to users' change requests and continuous integration, as promised by core Agile development technologies like DevOps [52], nevertheless, is not consistent with the intrinsic characteristics of blockchain harness. For example, the data that is generated and stored by smart contracts on distributed ledgers is almost impractical or too costly to modify by external users and developers. This limitation is mainly due to the immutable and append-only feature of blockchain environments in which smart contracts are executed [10],[12],[18],[19]. As will be discussed in Section 5.2.1, a conclusion from this matter is that Agile practices are thus useful at the early stage of BBS engineering to identify potential BBS requirements and smart contracts. They may not be always suitable for developing ever-changing business services. Rather, software teams should implement smart contracts for business services with minimum later upgrade needs at run-time and ensure the verification of smart contracts before their deployment on blockchain platforms. In this regard, [S23] suggests separating the system development lifecycle into two individual concurrent development endeavors, i.e., (i) smart contracts development and (ii) typical software development. Complementing this, developers should implement proxy/mediator components to enable interactions between these two endeavors to inform on-chain components of changes in off-chain components.

The benefits of agile development and blockchain are reciprocal. Whilst the above-mentioned studies narrow their focus on the contributions of agile practices for the engineering of BBS, other studies explore the ways in which blockchain features enhance and scale to the agile development. For example, blockchain can support the communication between product owner and developers in handling product acceptance test and certification via smart contracts in agile Scrum or Lean-Kanban process as discussed by Lenarduzzi et al., [53]. The benefit of blockchain to agile development falls out of the scope of this survey, although, it is an important potential research area.

*5.1.2 Model-driven BBS development*

The underlying principle of model-driven development (MDD) in software engineering is *abstraction* through which separation of concerns can be maintained between abstraction (domain models) and specification (implementation models) along with necessary transformation rules to move from abstraction to specification [54], [55]. The abstraction provided by models facilitates understanding of a given domain and automatic code generation while maintaining the traceability between models (system design) and code (system execution). A model can express the structure and behavior of domain concepts and define a set of implying statements about concepts and their relationships [56]. Numerous advantages have been shown for incorporating MDD into BBS engineering [S5],[S16],[S28],[S57],[S29],[S30],[S34],[S52]. Firstly, MDD-based development platforms, e.g., Ethereum and Hyperledger Fabric (See section 5.2.3), offer model transformation techniques and generate well-tested smart contract bytecodes to automatically detect vulnerabilities that may be exploited by attackers targeting blockchain platforms. Secondly, platform-agnostic models of smart contracts mitigate the issue of vendor lock-in to a specific blockchain platform as each platform may have its own programming language to implement smart contracts, which reduces the reusability and platform interoperability of smart contracts. If a developer needs to deploy the current smart contract, for example, written in Solidity language of Ethereum platform in Neo platform based on C# language, he/she needs to learn a new language and re-implement smart contracts. By following an MDD-based approach, once a smart contract is modeled, it can be transformed to multiple blockchain platforms. Thirdly, the traceability to requirements and consistency checking, for instance, from legal contract models to smart contract models and subsequently to smart contract code models is facilitated. Fourthly, from a human perspective, smart contract models are easier to understand than smart contract codes. Hence, they improve communication in/between blockchain engineers, domain experts, and stakeholders.

The abovementioned benefits can be embodied throughout a BBS development endeavor. An example of this application is the work by Jurgelaitis et al. [S5] that proposes an end-to-end development lifecycle. This work is inspired by the Model-Driven-Architecture (MDA) [57] approach, which defines three types of models that are gradually refined from high-level models to platform-specific ones based on vertical model transformation techniques. The authors use Computation Independent Models (CIMs) to represent BBS requirements and interaction models. CIMs are used to derive Platform Independent Models (PIMs) representing, in particular, high-level smart contracts that can be mapped to implementation details, e.g., Ethereum or Hyperledger. PIMs, give a broad view to software teams of what smart contracts should be in operation, help with a better selection of target blockchain platforms, specify transaction structures, participants, and consensus algorithms which are next reflected in Platform-Specific Models (PSMs). Model transformation rules are used to generate smart contracts for a chosen platform out of PSM models. In adopting MDA harness, software teams should identify what needs to be modelled and how it to be transformed. The discussion on the challenges of model transformation for the automatic generation of smart contracts is an important potential research area as discussed in the selected studies [S5],[S16],[S28],[S57],[S29],[S30],[S34],[S52]. However, it falls out of the scope of this survey.

*5.1.3 Architecture-based BBS development*

In software engineering, architectural thinking is a means for managing software development complexity by breaking down a large system into its components, interrelationships, existence rationale, and important properties. Software architecture models, such as *4+1 view* [58], enable the representation of different software architecture aspects such as logical, process, physical, and scenario. This enables examining if system quality factors are achievable [59]. The use of an architecture-centric approach in BBS development, as reported in [S3],[S4],[S11],[S12],[S33] is recommended for the purpose of complexity management. For instance, a logical view of the target BBS architecture, suggests decomposing BBS into different *layers* for separation of concerns where each layer is allocated to an individual team. The architectural view by Glaser et al. [S33] divides BBS development into three layers: (i) *presentation layer* where end-users interact with the BBS, (ii) *fabric layer* that fulfills basic services and the actual blockchain code base including communication layer, the public key infrastructure, data structures, and constructs for developing and execution of smart contract languages, and (iii) *application layer* that includes software components invoking basic services from the fabric layer. In a similar vein, [S4] and [S11] jointly define layers namely *application*, *trust*, *blockchain*, *transaction*, and *network*. Alternatively, developers can divide BBS architecture into *modules* to be designed and developed like (i) *REST API/front-end*, (ii) *blockchain Hyper-ledger*, and (iii) *back-end* *system*, where REST API/front-end allows end-users to interact with blockchain Hyper-ledger and back-end system [S3],[S12]. In section 5.2.2, we will further discuss the role of architecture-centric approach in BBS design phase.

*5.1.4  Pattern-based BBS development*

A software pattern represents a *general solution to a common recurring problem, from which a specific solution may be derived* [60]. Patterns, as a timeless piece of reusable knowledge, are useful to resolve repeated problems faced during development for multiple blockchain platforms. Zhang et al., [S35] contend that smart contracts have similar structures and concepts in object-oriented programming models such as inheritance, abstraction, and polymorphism in C++. Hence, many well-known software patterns can be employed to address the design challenges of smart contracts such as interoperability and reuse. Most patterns so far used for BBS, as discussed in Section 5.2.2, focus on the design phase. they are classified into three themes:

(i)  existing well-known design patterns in conventional software engineering that are applied to resolve BBS development challenges such as *interoperability patterns* proposed by Zhang et al. [S35];

emerging platform-agnostic patterns that are results of blockchain practitioners' experience gained while developing BBS in real-world case scenarios such as Solidity coding practices [S36], *smart contract design patterns* by Liu et al. [S37], *blockchain data migration patterns* by Bandara [S38], *gas consumption* [S36], [S39], and *interaction with non-BBS external systems* [S26],[S35];

(ii)  platform-specific patterns such as *Ethereum Solidity smart contract security design patterns* [S39], [S40]

For example, the collection of 15 patterns proposed by Liu [S37] has been identified and abstracted from different real-world BBS development scenarios. Bartoletti and Pompianu [S40] analyzed 811 Solidity smart contract source codes on the blockchain explorer (https://etherscan.io) and extracted design patterns related to transaction execution time management by a smart contract, i.e., *time-constraint patterns*, and encoding smart contract business logic execution to protect from critical operations, i.e., *math patterns*, and so on.

A unique and typical challenge of smart contract design is the gas cost consumption for the execution of transactions on blockchain platforms. Each transaction that is executed by smart contract will be charged some amount of gas, so the more intensive the computations are, the more gas fee should be paid for. The execution of a smart contract may be failed if the transaction is excessive in gas consumption. On the other hand, an accurate estimation of gas consumption by a smart contract may not be easy. From this point of view, design patterns and recommendations by [S36], [S39] can aid software teams for smart contract design, with a view to save gas.

From these pattern collections, it appears that, although software patterns have gone some length to effectively address the problems of BBS design phase, their application to requirements elicitation and analysis is not yet explored in practice. Patterns related to requirements engineering enable software teams to capture right requirements, whose realization via BBS creates added values, which an important avenue for future research. Another matter related to the deficiency of literature in adopting patterns for blockchain engineering is that almost all proposed patterns contribute to better BBS design. Despite their usefulness, there is a lack of patterns for specifying BBS development processes, which technically are referred to as *process patterns*, a term coined by Coplien in his landmark paper in 1994 [60]. Process patterns, for example, as shown in [61],[62], are the result of applying abstraction to recurring software development processes, thereby providing a foundation for constructing a bespoke software development process through the composition of appropriate process pattern instances. Applied in BBS engineering, process patterns can be an invaluable source of insight for blockchain researchers and practitioners because they can normally reflect the state of the BBS development processes and are based on recurrent series of actions and well-established concepts.

*5.1.5  Ontology-based BBS development*

Ontologies help reduce conceptual ambiguities and inconsistencies in a particular domain while enabling value-creation capabilities [63]. The application of ontologies becomes important to facilitate knowledge interoperability among stakeholders. Ontology-based system development is initiated by identifying concepts/classes in a domain and is followed by assigning properties for each concept and defining domain constraints and relationships among these concepts that need to be verified if constraints are violated by these properties [64]. In view of the identified studies, the purpose of ontology-based BBS development is for training [S41], identification, design, and test of smart contracts [S42],[S43], and improving discoverability of smart contract services at run-time [S44]. Kruijff and Weigand [S41] discuss the support of ontologies for BBS development by providing essential concepts related to an operational BBS. Their proposed ontology can be viewed as training material for software teams who might not be familiar with the key concepts of blockchain. This pioneering effort, in turn, has motivated others to explore alternative applications of ontologies to support the development activities of BBS, in particular the design phase, as will be delineated in Section 5.2.

## 5.2 RQ2: What are the key emerging and challenging tasks in developing BBS?

BBS engineering aims at achieving business outcomes such as lower transaction costs, reduced dependency on trusted third parties, regulation compliance and certification, and secure data access. To this end, a series of tasks are needed to accomplish these outcomes. We identified these tasks from our primary studies as depicted in Figure 4. The tasks are grouped into four typical SDLC phases.

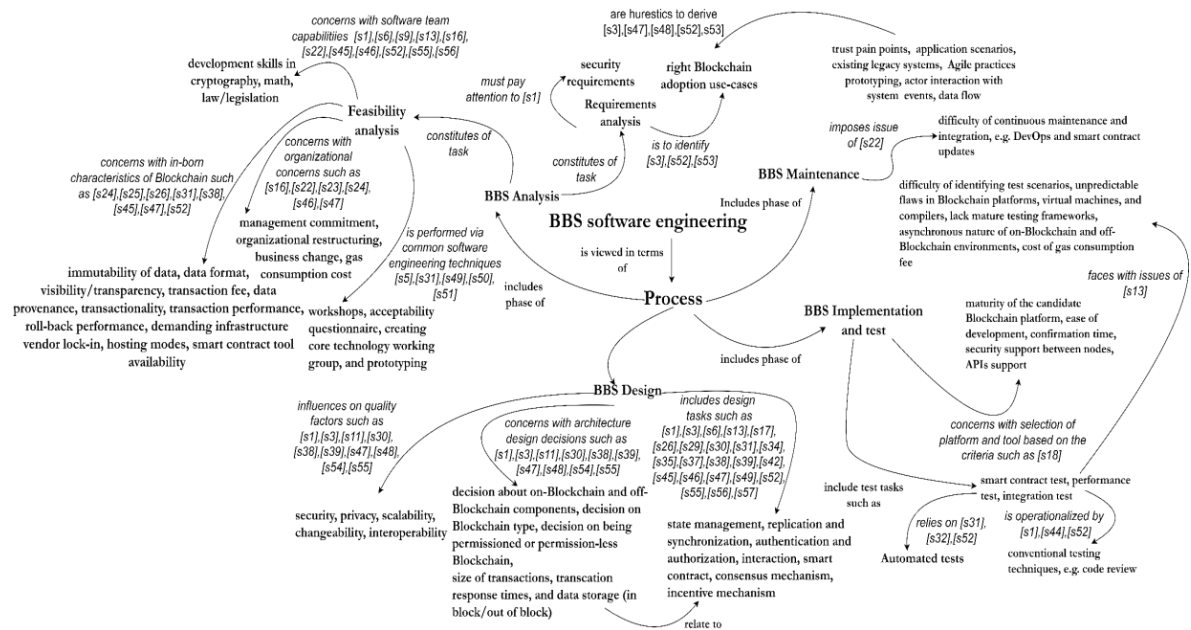

Fig. 4. *Process* conceptual map

### 5.2.1 BBS analysis

The analysis phase encompasses tasks related to requirements elicitation and verification. It establishes a link between stakeholders' needs and desired features of the target BBS. Note that the tasks *feasibility analysis of BBS* and *requirements analysis of BBS* in our framework are by no means a complete description of the comprehensive analysis phase of BBS engineering. They highlight some distinctions of BBS development from a non-chain software engineering perspective. We elaborate on major challenges related to this phase as follows.

*Feasibility analysis of BBS.* BBS is not merely a technical shift and a subject of regulation but also a substantial change in the way all stakeholders and systems hereinafter will operate, provide services to users, and be maintained. A BBS might be at the same time disruptive to some stakeholders and sustaining to others. Hence, the first essential task in the development process is to examine if blockchain technology enablement is a suitable option to embed into organizational business processes and to come up with an outline plan for the next phases. The most recent literature enumerates important *situation factors* (or *competing factors*) that may inhibit BBS development process and thus should be examined in the *feasibility analysis* task as listed in Table 3.

Table 3. Sample list of *situational factors* for consideration in *feasibility analysis of BBS* task

| Category | Situational factor | Possible values for the factor | References |
|---|---|---|---|
| Organization | Management commitment | Low, medium, high | [S16],[S24],[S46],[S47] |
|  | Organizational restructuring | Hierarchical, functional, horizontal, divisional, matrix, team-based, network | [S16],[S24],[S46],[S47] |
|  | Business change | Rarely, occasionally, frequent, always | [S22], [S23], [S24] |
| Software team | Energy and gas consumption cost | Low, medium, high | [S1] |
|  | Development skills in cryptography, law/legislation | Low, medium, high | [S1],[S6],[S9],[S13],[S16], [S22],[S45],[S46],[S52], [S55],[S56] |
| Technical | Immutability of data | Stable, volatile | [S24] |
| | Data format | Homogeneous, mixture, substance | [S38] |
| | Visibility/transparency | Public, private, protected, package private | [S24],[S45] |
| | Transaction fee | Free, low cost, variable, expensive | [S38] |
| | Data provenance | Source, temporal, meta-data result | [S38] |
| | Transactionality | Low, medium, high | [S24] |
| | transaction performance | Critical, fairly important, low, moderate | [S24],[S25],[S26],[S47] |
| | Roll-back performance | Critical, fairly important, low, moderate | [S24] |
| | Demanding infrastructure | None, emergent, operating, matured | [S24] |
| | Vendor lock-in | Impossible, unlikely, even chance, certain | [S48] |
| | Hosting modes | Cloud, dedicated servers, virtual private, | [S38] |
| | Smart contract tool availability | Yes, no | [S31], [S52] |

—*Immutability* achieved through cryptographic algorithms, can be a cost barrier to a BBS project [S24]. That is, if the need for the data persistency, immutability, and records' longevity is not greatly important, BBS may be an expensive choice, as an implementation effort will be required in the subsequent phases, in comparison to other data persistency mechanisms such as normative database management systems (e.g., Relational Database Management Systems). Stakeholders need to decide if immutability is a critical need to be addressed by a BBS development instance.

—*Visibility and transparency* indicate that BBS enables all participants in the network to observe the chain of blocks, regardless of privacy and anonymity applied to the data, download blocks, and check them against contract rules. For example, participants can validate if there is a certain threshold for ordering items from a specific supplier, automatically via smart contracts. Visibility to external actors and systems is a situational factor that should be taken into account in the BBS project feasibility analysis task [S3],[S24].

—*Organization restructuring* means that BBS may raise complexities and changes in intermediary trusted legacy systems, cooperating organizations, boundaries, and layered based organizational structure for the coordination and enforcement of business legislations and privacy rules. This often results in a decentralized structure comprising of peers working according to fully transparent rules which are written in smart contracts [S16], [S24],[S46],[S47]. Hence, smart contracts should be adopted only if all parties agree to perform transactions on distributed ledgers. Findings in the research by Du et al., [S47] highlight that an incremental approach takes precedence over a revolutionary approach for organizations to develop BBS to accumulate tangible outcomes of addressing stakeholders' concerns. Moreover, BBS presents a risk of software maintenance team downsizing. That is, as BBS increases the automation of decision making such as documentation, monitoring, verification, and auditing reports that are traditionally manually performed, staff associated with these activities will no longer be needed. This may raise the resistance of organizational staff to BBS development.

—*Transactionality* of organizational business services needs to be evaluated before converting them to smart contracts. If the type of business services is not concerned with verifying and enforcing credible processing transactions and data exchange between permitted participants, then BBS development is not worth it. For example, blockchain may not make an applicable case for simulation systems where the credibility of experimental data is not concerned. In such a situation, Scriber et al., [S24] suggest that non-transactional services may indirectly benefit from BBS via intermediator/proxy software components – depending on the definition of use-case scenario- where the intermediator sends/receives transactions between BBS and non-transactional services.

—*Operational performance* relates to the result of integrating existing legacy software systems with BBS. This evaluation should be based on factors such as acceptable latency and level of transaction verification. Unlike non-BBS systems, in which a transaction verification can be determined instantly, BBS harness is attributed by limited operational performance due to the latency of associated Proof of Work (PoW) algorithms to reach consensus and interoperability amongst chains/ledgers as pinpointed by findings of [S24], [S25], [S26], [S47]. A node might accept a transaction, but wait for an extended period, based on the defined consensus model, for verifying, sharing, accepting, and encoding a transaction into the chain of blocks. If the target BBS is expected for a real-time response or specific performance, i.e., processing millions of transactions continuously, BBS will affect the overall performance and scalability [3]. To alleviate this issue, Du et al., [S47] suggest that software teams should, in collaboration with stakeholders, identify an essential dataset from business transactions for storing on blockchain and keep non-essential transactional data on legacy systems. Furthermore, data and transaction security enforced by cryptography and PoW algorithms affect expected BBS performance. In particular, for an IoT-based BBS, constrained devices might not be able to perform more advanced cryptographic techniques due to bandwidth limitations, battery capacity, memory, processing power, and heat management issues [S24].

—*Demanding infrastructure (or technical maturity)* refers to providing support for the high degree of transaction processing, computational power to perform cryptography algorithms, storage size, and high bandwidth for distributed nodes, which are essential to deploy and maintain BBS. Organisations in poor regions or rural areas may not be able to afford these costs if they are to set up BBS on their own infrastructure [S24].

There are other concerns that cause BBS to turn out ill-suited and far-reaching consequences in later phases that will be costly to rectify. These concerns, mainly related to the current technical limitations, are *gas consumption cost* (computational effort required to execute operations by smart contract) [S1], *development skills in cryptography* [S1],[S6],[S9],[S13],[S16],[S22],[S45],[S46],[S52],[S55],[S56], *business change* [S22],[S23],[S24], *delayed rollback for cancelled transactions after verification* [S24], *differences in data format*, *hosting modes*, *transaction fees*, *data provenance* [S38], *vendor lock-in/platform* [S48], and *tool availability for smart contracts test* [S31],[S52].

Software teams can embed the existing conventional techniques into the development process to conduct the *feasibility analysis of BBS* task such as conducting *workshops* [S31], [S49], *acceptability questionnaire* [S5], creating *core technology working group* [S50], and *prototyping* [S51] to evaluate if BBS simplifies the business processes and is an appropriate fit. It should be noted that the abovementioned factors may influence each other and hence should be investigated before moving to further phases of the BBS development process.

*Requirements analysis of BBS.* A highlighted challenge of requirements analysis in the context of BBS development is to identify right and cogent use-cases. Plansky et al., [S50] propose a four-step technique starting with the compilation of a pilot project that all stakeholders believe a distributed ledger can contribute to strong prospects. Existing pain points such as the lack of trust in the current collaborative business processes, conformance checks, intermediary delays, and areas of user dissatisfaction should be identified. Explicit hypotheses on how BBS improves these issues in terms of decreasing a certain level of organizational cost in a specified time period are defined in the second step. This is followed by developing the BBS prototype, for example, a smart contract, to attest hypotheses in the third step. If the prototype gets a clear sense of hypotheses acceptance, a project plan aiming at a few long-term goals e.g., increased revenue, better compliance, cost reductions, is set for the scaling up the prototype in a measurable way. In the same vein, Fridgen et al., [S52] propose a technique with application examples in five idiosyncratic industrial cases. The technique consists of the following steps: (i) understanding blockchain technology to get a conceptual and technical foundation, (ii) deriving BBS application scenarios, (iii) analyzing existing legacy systems, (iv) documenting application scenarios, and (v) and developing and evaluating the prototype. Moreover, the requirements analysis in the proposed development process by Marchesi et al., [S3] relies on Agile practices and includes the following steps: (i) specifying blockchain adoption goals visible to stakeholders, (ii) identification of actors interacting with target BBS, e.g., human, external system, and IoT devices, (iii) determining the trust requirements between actors and BBS components, and (iv) documenting requirements in the form of user stories and expected features. Alternatively, the approach of Du et al., [S47] underlines the role of blockchain as a new *automated trust-building machine*. Accordingly, software teams should identify user inputs, manual verification, and reconciliation in transactions as candidate areas for converting them to smart contracts that create self-executing transactions through if-then conditions. A particular group of requirements that need more attention is related to security, which should be captured at the early stage of the development and be addressed in the design and test phases [S1]. Other than that, the literature suggests further techniques for requirements identification such as *events occurrence identification* [S48] and *data-driven* [S53]. They specify data elements of a problem domain and uncover data flow that is needed to be transformed and managed by smart contracts.

### 5.2.2 BBS design

The basic assumption in the design phase is that it leverages the output produced by the analysis phase (Section 5.2.1). The design phase identifies features to be exhibited and requirements to be satisfied by a specific architectural blueprint of target BBS. According to the identified papers, discussion on the design phase is organized into three essential architectural design decisions including (i) *decision on-chain and off-chain components,* (ii) *decision on blockchain type,* and (iii) *decision on being permission-based or permission-less blockchain*. In addition, the phase defines seven key design tasks namely: (i) *state management design*, (ii) *replication and synchronization design*, (iii) *authentication and authorization design*, (iv) *interaction design*, (v) *smart contract design*, (vi) *consensus mechanism design*, and (vii) *incentive mechanism design*. Both key architectural design decisions and tasks should be centered on the common system quality factors such as privacy, security, energy efficiency, scalability, and interoperability. They are realized in the next phase by implementing new software and hardware BBS components. The following elaborate on these key architectural design decisions and tasks.

*Key architectural design decisions.* The architecture of a BBS can be viewed as the set of interrelated design decisions that influence the quality factor of the overall BBS functionality. Thus, decisions are indispensable to the success of a BBS development endeavor. These decisions include:

—*Decision about on-chain and off-chain components.* This determines functionalities that should be kept on-premises and functionalities that can be deployed in a blockchain network [S54]. Such segregation between on-chain and off-chain components, as defined in [S1],[S11],[S30],[S38],[S39], [S54], facilitates the separation of concerns and the manageability of implementation and maintenance of a BBS. On-chain components are smart contracts that are deployed and executed on blockchain platforms whereas off-chain components host off-line data and application business logic. Off-chain components interact with on-chain components via transactions. The interaction between these two groups of components is enabled via *mediator components* [S30] or *gateway connector components* [S54] that provide services for communication (transferring data), coordination (transferring control), conversion (adjusting unmatched interactions), and facilitation (optimizing interactions), as depicted in Figure 5.

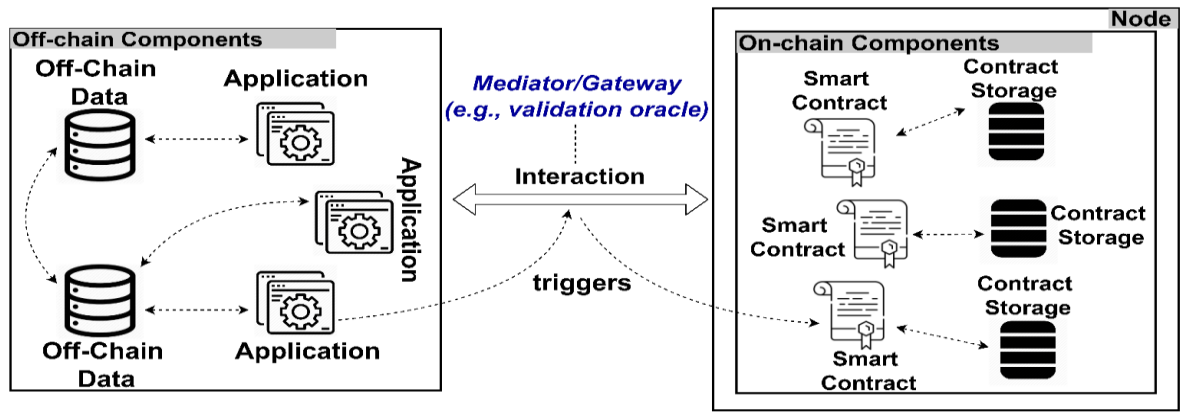

Fig. 5. A BBS architecture representation based on off-chain and on-chain components

—*Decision on blockchain type.* The type of blockchain network - public, private, and consortium- leads to making a trade-off between quality factors such as privacy, performance, traceability, and transparency. Unlike private and consortium blockchains where a single or a limited number of organizations authorize and control transaction validations, a public blockchain may include a large number of nodes. All committed transactions in a chain of blocks are visible, downloadable, and verifiable to all nodes in the blockchain network. Whilst this enables better information transparency and auditability, on the other hand, such transparency in public networks is at the expense of information privacy and scalability degradation for a large chain of blocks. Another example of trade-off architecture decision-making situation is that both private and consortium types offer partial immutability because participants can agree within the network to change, update, or delete a previous record. Despite the immutability constraints, private and consortium blockchain have significantly higher transaction throughput compared to a public blockchain as well as cheaper set-up cost due to less required resources. Lin et al. [S1] recommend that public blockchain should limit the number of smart contracts and writing of permanent data due to the high costs such as computational power, gas consumption, and data capacity involved in running computations and recording in the permission-less network. Therefore, software teams should not move all computations and data to the blockchain. A common recommendation, as jointly pointed by [S1], [S3], [S47], [S54], is to keep the big and private raw data off-chain whilst the tamper-resistant, transparent, and traceable data reside on-chain. For instance, the hash of personal data can be stored on-chain and the raw data are saved to off-chain to provide better scalability and higher efficiency.

—*Decision on being permission-based or permission-less blockchain.* This concerns the extent to which a participant plays in a blockchain network. The authority of permission management is granted by a permission-based blockchain network. Permission definition ranges from joining the network, transaction submission, transaction validation, asset creation to mining operations. For example, a participant node that is allowed to join the network has the read permission on the recoded transactions in blockchain. In a permission-less architecture, e.g., Bitcoin and Ethereum, a participant can be either pseudonymous or anonymous. In a permission-based architecture, e.g., Ripple and Eris Industries, the identity of participants is identifiable, which is similar to conventional commercial bank accounts. Hence, in permissioned architecture, participants need a legal identity to validate transactions. Both public and permission-less blockchain architecture do not guarantee data security and privacy. In other words, any node can join the network and visit all data records of participants. Since legal contracts that are executed by smart contracts will be accessible by participants in the public blockchain, software teams may be reluctant to use BBS or they need to select a permissioned blockchain architecture to grant or revoke permissions to participants. The permissioned blockchain can enforce access control to participant nodes. Whilst developers might want to encrypt transaction data on public and permission-less blockchain through cryptography algorithms to alleviate the privacy issue, on the other hand, a side effect of the encryption is the overall BBS performance and throughput degradation due to processing time needed for the data encryption/decryption. Hence, in the selection between permissioned or permission-less architecture for BBS, software teams should make the trade-offs between factors such as security, transaction processing rate, cost, censorship, and reversibility [S55].

There are other design decisions listed by Arthur [S48] that influence architecture design, such as the *size of transactions*, *speed of response times*, and *data storage (in block/out of block)*.

In the following, we review seven essential design tasks that are frequently highlighted by our selected papers. We view the necessity of these tasks in relation to the layers of blockchain architecture as shown in Table 4. Based on the studies [S3], [S11], [S12], [S31], [S33], the following distinct layers of BBS are notable:

—*Application layer* packages existing legacy systems or new software applications interacting with smart contracts;
—*Smart contract layer* includes various scripts and algorithms that execute certain operations on data, memory, business process, and assets stored in blockchain network once pre-conditions are met;
—*Incentive layer* defines mechanisms for economic rewards in blockchain network to motivate participants to continue their effort for new block validation once it is created;
—*Consensus layer* contains possible consensus algorithms to reach transaction validity once it is added to the blockchain;
—*Network layer* specifies mechanisms of distributed networking for data forwarding, routing, and verification;
—*Data layer* stores the chain of data blocks via related mechanisms such as asymmetric encryption, time-stamping, and hash algorithms; and
—*Physical layer* includes hardware components such as servers, networks, and IoT devices on which BBS components are deployed.

Layering in BBS helps maintain logical separation of concerns (a.k.a. functional or operational slicing of a system), where data or control flow moves from top to bottom or left to right layers, typically indicated with a directional arrow as in Figure 6. Strict enforcement of layering requires that the top or left layer must only transfer data or control to the layer immediately to its bottom or right layer, however, for practical reasons, the hierarchy can be skipped. For example, as in Figure 6, a BBS that does not follow incentivization and consensus mechanisms can skip the incentive and/or consensus layer to move directly to the network layer for further processing.

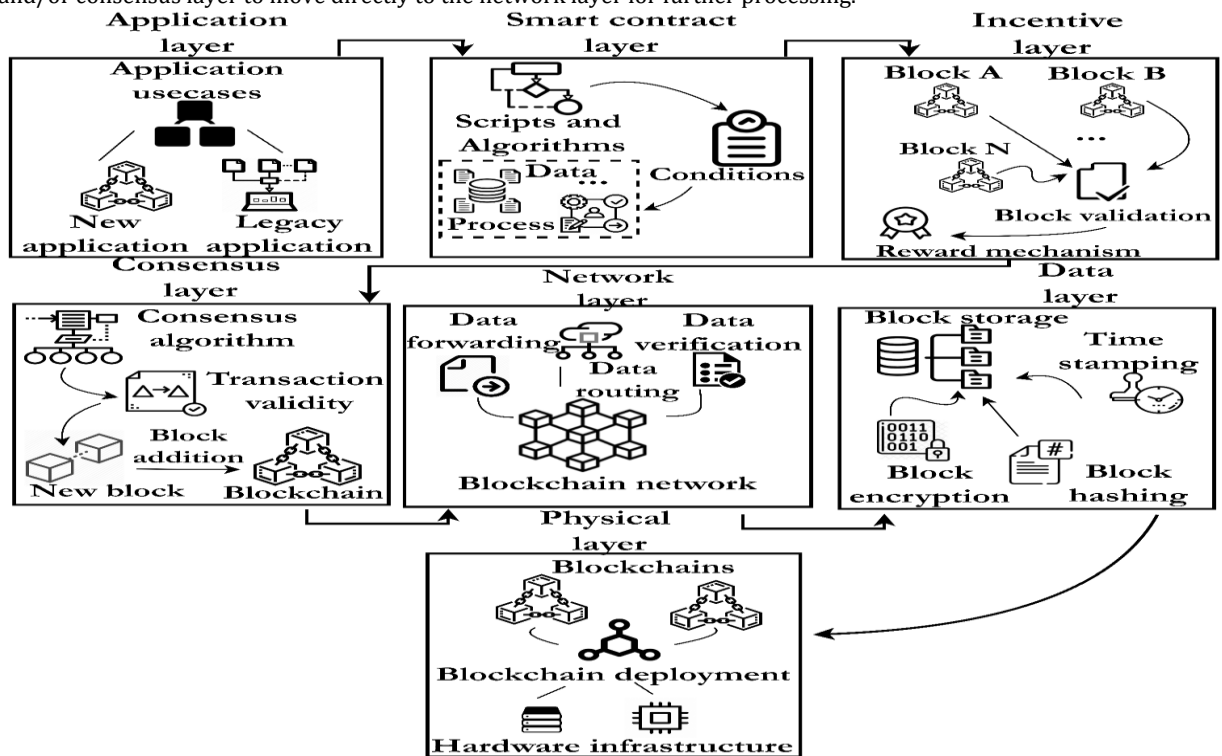

Fig. 6. Architecture layers in BBS

In performing the design tasks, software teams face trade-offs among many quality factors across different layers of BBS architecture. These are explained along with illustrative examples in the following.

Table 4. Design tasks based on their correspondence to blockchain architecture layers

| Task \ Layer | State management | Replication and synchronization | Authentication and authorization | Interaction | Smart contract | Consensus mechanism | Incentive mechanism |
|---|---|---|---|---|---|---|---|
| Application | × | × | √ | √ | × | × | × |
| Smart contract | √ | × | √ | √ | √ | × | × |
| Incentive | × | × | × | √ | × | × | √ |
| Consensus | × | √ | √ | √ | × | √ | × |
| Network | × | × | √ | √ | × | × | × |
| Data | √ | √ | √ | √ | × | × | × |
| Physical | × | × | √ | √ | × | × | × |

*State management design.* In an on-chain environment, a *state* is a snapshot of the distributed ledger at a specific time. The state changes as a result of creating, updating, or deleting blocks. In such a self-contained environment, a ledger keeps a full record of state transitions in a timestamp-sequenced, immutable, and tamper-resistant way. In an off-chain environment, however, external software components (or off-chain components) do not keep these states to perform their transactions. This means the state of external systems is not accessible by on-chain components, for example, smart contracts (Figure 5). This can be problematic for the validation of cross-domain transactions that may depend on the states of both on-chain and off-chain environments. To handle this challenge, some useful design patterns are proposed by Xu et al., [S26] and Zhang et al. [S35]. These are concerned with *interaction with the external world* and resolving the interaction issues between on-chain and off-chain components [S26], [S35]. The patterns are suggested to design state management components and they are based on the notion of (i) *Oracle*, i.e., storing the state of external systems into the blockchain execution environment and (ii) *reverse Oracle*, i.e., creating an interface between external systems and smart contracts for supplying data and check conditions, and (iii) *validation oracle* i.e., a state coordination mechanism to check conditions that cannot be represented in an on-chain environment. If a transaction validation depends on external states, the validation oracle is asked to validate and sign the transaction. The validation oracle mechanism can be performed by automatically or by a human arbitrator injecting external states into blockchain via regularly updating the values of external application states within the smart contract storage. Alternatively, Luu et al. [S17] propose a technique in which a transaction $T$ defines a guard condition $g$. The current state should satisfy $g$ for the execution of $T$ to proceed. If $g$ is not satisfied, the transaction is simply dropped. For transactions that do not provide $g$, it is simply considered $g \equiv$ true. This solution guarantees that either the sender gets the expected output or the transaction fails. Nevertheless, software teams should be aware that all these ways affect the execution performance of transactions and cause a delay due to the need to make sure the consistency of states between on-chain and off-chain components by *validation oracle* mechanism [S55].

*Replication and synchronization design.* There are differences in the design of data replication and synchronization between conventional software systems and BBS [S55]. In the non-BBS context, software teams adopt common data partitioning and replicating mechanisms such as master-slave or multi-master replication to improve the throughput, sustainability, and latency of a software system. In contrast, in BBS harness, each data block is duplicated over network nodes, which causes unwanted increase in latency and the reduction of BBS throughput. Moreover, for the data synchronization design in conventional software architecture design, there are well-known mechanisms such as 2-Phase Commit and Paxos to keep data replication synchronized. The synchronization mechanism in BBS is based on reaching consensus among network nodes to perform correct operations regardless of faulty components. Additionally, consistency control on performing transactions in BBS is based on the defined terms and rules in their associated smart contracts. The choice of replication and synchronization mechanisms [12],[18],[19],[37], will be, unavoidably, influencing final BBS quality factors of throughput, sustainability, and latency.

*Authentication and authorization design.* Specific mechanisms should be designed to enable the identity detection of an entity in a blockchain network, which either manages data for operational/business purposes or provides services to participants, via cryptography techniques such as digital signatures, homomorphic encryption, and multi-party computation [12],[18],[19]. In particular, for a permissioned blockchain, an additional access control layer should be defined via Certificate Authority (CA) and a Membership Service Provider (MSP). Hebert et al., [S49] list a comprehensive set of security risks including fifteen general categories of risks from key management to privacy, consensus, interoperability, and legal concerns. To tackle these risks, [S1] and [S3], jointly, define a set of security design principles that should be realized via implementing new components in BBS architecture. The following sub-tasks are recommended:

—*Log design*. BBS reflects historical records and the usage of data by any entities such as who, why, when, what and how, that all should be logged in the ledger in a verifiable way.
—*Policy design for data usage*. The policy indicates rules for topics such as access rights, permissions, and conditions.
—*Off-Chain data storage design*. Storing sensitive data directly in blockchain, regardless of encryption techniques applied, can be at the risk of privacy leakage. Thus, the sensitive data should be stored in off-chain components such as conventional database management systems (e.g., Oracle or MongoDB), cloud storages (e.g., S3, AWS or Azure), and distributed storages (e.g., IPFS or Storj). The reference to these data should be stored in the blockchain. Such a reference can be a hash, a connection string, an absolute path, or an identifier referring to a dataset.
—*Circuit breaker design*. This indicates necessary emergency termination functions in smart contracts that are triggered in the case of bugs or the occurrence of abnormal behaviour by an entity.
—*Interaction design*. This task is to specify interdependencies, integration pain points, and the way BBS components can communicate, exchange, and use each other's services. It helps characterize the overall architecture and the position of BBS in the operational environment. A key concern during the interaction design task is the interoperability factor across the BBS architecture layers (Figure 6). This is due to reasons like (i) switching BBS between single or multiple blockchain platforms, (ii) BBS integration with existing legacy software applications, and (iii) multiple types of BBS that support the completion of a cross-domain transaction within networks, which involve computational resources owned by different entities, (iv) changing hardware components on which BBS instance is running, and (vi) changing a smart contract to a new one or combing several smart contracts into a single one. For instance, some smart contracts that are upgraded to new Ethereum Virtual Machine (EVM) with the aim of strengthening the security and optimizing performance, may raise interoperability issues – in the scope of the platform versions, smart contract languages, and APIs. Hardjono et al., [S56] investigate the interoperability issues in BBS architecture design and propose the following root cause analysis to identify interoperability pain points: (i) *minimal assumption*, i.e., identifying the unit of transaction that is semantically recognizable for multiple BBS, (ii) *degrees of permissionability*, i.e., identifying transactions that are processed by both permissioned network and external permission-less networks (foreign domains), and (iii) *participants*, i.e., identifying network's nodes that are involved in transaction processing. In addressing the interoperability pain points for cross-domain transactions, Hardjono et al., [S56] rely on the notion of peering agreement in autonomous systems and recommend steps of designing (i) semantic compatibility that is required for multiple BBS, (ii) data exchange and communication protocols, (iii) mediators, delegations, and technical-trust mechanisms, and (iv) agreements such as level of services, fees, penalties, liabilities, and warranties.

Recall from Section 5.1, software patterns can be used to resolve interoperability challenges between on-chain and off-chain components. Zhang et al., [S35] exploit the application of software design patterns to address BBS interoperability design challenges and choose e-health software applications as a representative use case. The application of four sample software patterns, namely *abstract factory*, *flyweight*, *proxy*, and *publisher-subscriber* is appraised. For example, to abstract away the implementation detail of data storage of a smart contract, the proxy pattern is suggested that enables seamless interaction between BBS components while still supporting variations in smart contract's data storage options. Moreover, the abstract factory pattern provides the creation services to instantiate generic smart contracts to domain-specific ones. Heterogeneity of blockchain platforms and legacy systems and the need for the support of data migration from on-premise to blockchain platforms or migration between multiple blockchain platforms have been taken into account from a software pattern-oriented view. To this end, Bandara et al., [S38] have proposed *blockchain data migration patterns* such as *state extraction pattern*, *state transformation patterns*, *state and transaction load patterns*, and *safe data management*.
—*Smart contract design*. The aim of this task is to (i) analyze existing textual legal contracts, (ii) identify and convert semantic clauses and rules that are expressed as smart contracts, and (iii) transform smart contracts into software codes. Derived from our selected papers, we divide the task of smart contract design into two sub-tasks of *(preliminary) high-level design* and *(detailed) low-level design* as shown in Figure 7.

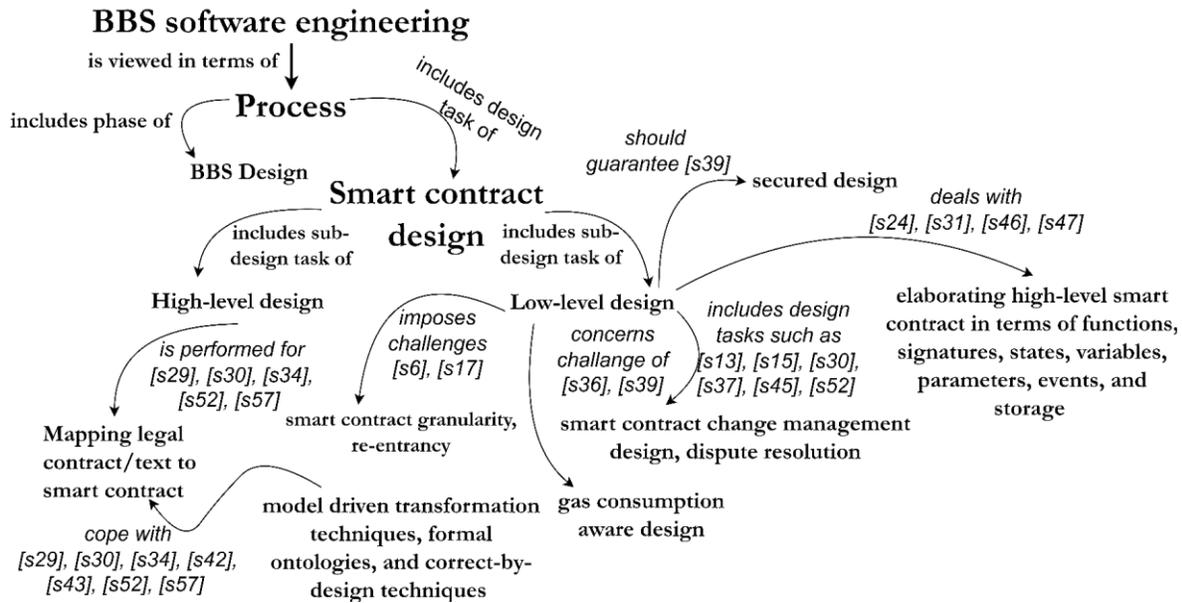

Fig. 7. *Smart contract design* conceptual map

The *high-level design* sub-task is concerned with crafting an initial structure for smart contracts by translating clauses of legal (textual) contracts, business transaction logics, or backend codes into smart contract codes, e.g., Ethereum Solidity scripts that are executable on blockchain platforms. A key challenge in smart contract design is how to map legal contracts to executable software code in a correct manner. There are important differences between the content of a legal contract expressed in the natural language and a corresponding smart contract code. Some prose of a legal contract may not have equivalent expressions in a smart contract and vice versa. Moreover, a legal contract may contain certain terms like a reasonable excuse, unintended delay, exempted penalty, and so on that are deemed implicit in a natural language. However, they may have no explicit meaning at the level of smart contract code. Hence, in transforming contract documents and legal prose to equivalent smart contracts, software teams encounter ambiguities and different interpretations that can be reflected in smart contract code.

As briefly introduced in Section 5.1.1, using ontologies and formal/mathematic expressions to describe legal text can potentially reduce the transformation errors in legal contract to smart contract mapping and ensure legal prose that are expected to be met. However, this needs training effort and acceptance by all BBS project communities (software teams and stakeholders) to use formal constructs to describe legal contracts. Domain ontologies are suggested to enhance the identification of target smart contracts including their functions' specifications, relationships, automated inference, and verification. Kim and Laskowski in [S42] use first-order logic to express the ontology of smart contracts that later execute a provenance trace and to enforce traceability constraints on transaction execution on blockchain. Following this introductory work, Choudhury et al. [S43] introduce a framework that uses semantic rules to encode BBS functional requirements and to leverage the structure of abstract syntax trees for the automatic generation of smart contracts. Also, ontology design techniques [63] enable developers to elicit potential network participants in business processes. They also help identify participants' relationships, conditions, and constraints in terms of resources exchanged between themselves which can be, accordingly, represented by smart contracts. Furthermore, ontologies are used to improve the discoverability of smart contract services that are offered by a deployed BBS. The notion of *semantic smart contract* is proposed by Baqa et al. [S44] uses RESTful semantic web to enable indexing, browsing, and annotating of smart contracts. As such, annotated smart contracts deploying on distributed ledgers can be treated as linked data for performing queries by domain-specific terms, which increases the discoverability of public smart contracts.

In addition to the application of ontology design techniques in the smart contract design, recall from Section 5.1.2, MDD can play a key role in the design phase of BBS engineering, in particular, in reducing mapping complexity, i.e. enforcing control-flow and conformance to normative business processes via smart contracts [S29],[S30]. Frantz and Nowostawski [S57], propose a semi-automated technique for transforming human-readable behavior specifications to a smart contract format. They borrow the idea of *institutional analysis* from the social and economic field in order to analyze and decompose the functions of human institutions into a set of declarative rule-based statements. Statements are expressed via core elements such as attributes, actions, outcomes, conditions, and consequences associated with non-conformance. The mapping of core constructs, which is automated via a domain-specific language (DSL) such as in Scala, provides a foundation to transform rule-based statements into smart contracts. An alternative design technique is proposed by Clack et al., [S52]. It is based on the idea of separation of concerns through which three levels of abstractions for a smart contract are defined: *legal contracts*, *smart contract templates*, and *smart contract codes*. In this layered-based contract design, a legal contract includes text (a linear sequence of sentences, hierarchical structure) in the natural language, lists/tables, cross-references, and redacted text (privileged and proprietary text). A smart contract template is an intermediate and electronic representation of the legal contract documents. A template contains different parameters such as identity, data types, and values. It is developed by standard bodies and uses parameters in legal documentations to connect it to corresponding codes. All stakeholders should have agreements on these parameters and their values as they directly influence business relationship as well as the way codes in smart contracts will operate. While a smart contract template tries to formalize actions, legal rights, and enforceable obligations that accrue to different actors and often are context-specific, a smart contract is an automated version of the smart legal template by a software script. This harness forms the layers of transformation which can be performed in a semi-automatic way using appropriate tools. A similar technique for high-level smart contract design, named *correct-by-design development*, is proposed by Mavridou et al. [S34]. The technique is based on an end-to-end formal model-based verification. It describes the security requirements via abstract transition models along with semantic operations and reasoning at the abstract model level to examine if the behavior of the contract models satisfies these requirements. The formal models are then transformed to Solidity code bytes for execution on Ethereum.

Smart contracts should provide mechanisms to guarantee the security and privacy of sensitive data access and process. Yet security challenges related to smart contract design are multitude. Although it is not in the scope of our survey to cover all the security-associated challenges of smart contracts, there is a rich collection of patterns, proposed by Wohrer et al. [S39] that deal with the security concerns in the smart contract design. Software teams can apply patterns such as *checks-effects-interaction*, *speed bump*, *rate limit*, *mutex*, and *balance limit* for safe and reliable execution of smart contracts.

*Low-level design*. Concerning the elaboration of high-level smart contract definition on functions' signature, parameters, contract storage, events, and state variables, the reviewed papers recommend performing the following alternative analysis:
—analyzing actors who interact with smart contracts ([S31], [S46], [S47]);
—analyzing interaction points and message flows between smart contracts and off-chain components, e.g., smart contracts and legacy systems [S46];
—analyzing events where a BBS is viewed as a stimulus-response machine reacting to smart contracts and off-chain components, e.g., smart contracts and legacy systems [S46];
—analyzing external behaviors of a BBS to identify states that a smart contract can enter or trigger events [S31], [S46];
—identifying the cancellation of transactions and delayed rollback [S24]; and
—specifying required libraries, data structure, inheritance structure, and interfaces for smart contracts [S31].

The low-level smart contract design also raises two common challenges that software teams should be aware of:
—*Granularity*. The granularity of a smart contract is defined in terms of the number/size of required processing functions in smart contracts to enforce rules and control the execution of business transactions [S6]. For large-scale transactions, an individual smart contract negatively affects power consumption, gas usage, storage space, and performance. For instance, Ethereum charges a gas fee for smart contract execution. Likewise, smart contracts are performed sequentially on blockchain platforms, which reduces the overall performance of BBS if there are many smart contracts that are bounded to a business process. The parallel execution of smart contracts is a resolution technique to address this issue. However, a detailed discussion on the concurrent smart contract models [12],[18],[19],[37] falls out of the scope of this survey.
—*Re-entrancy*. This challenge occurs in an interaction between two smart contracts when the control of smart contract execution is taken over from one to another during the execution of cross-functional transactions. Once a smart contract calls another smart contract, its current execution is suspended until the other call is finished. The intermediate status of the first smart contract may not be ready or correct to use by other BBS components. This, for example, can be a critical issue associated with the money withdrawal function. Smart contracts should define discrete, atomic, and sequential functions in a way that when a function is called, it cannot be re-entered by an external call before its current execution is entirely ended [S17].

Apart from the above discussion on high-level and low-level smart contract design, the literature draws attention to sub-design tasks related to smart contracts:

—*Change management design*. As mentioned earlier, legal contracts may include complex conditions and terms that are updated based on mutually agreed upcoming changes by parties. On the other hand, once a transaction is generated and stored by smart contracts on blockchain database, it is infeasible to modify this transaction as the blockchain is merely an append-only that supports the creation of a new transaction to add to the chain of blocks instead of updating already existing blocks. This is realized via defining change management function and control variables in smart contracts. This is in contradiction with evolving requirements of business environments. Unforeseen changes in legal prose and rules frequently occur after a contract document is finalized and signed by parties due to reasons such as a change in business goals or valid rescission by courts (no longer a cause of action for agreement breach) that yield in changing smart contract parameters and functions to meet new requirements. This is crucial to address change requests in smart contracts and to assess the change impact and propagation for keeping consistency between stakeholders' requirements and the design of smart contracts. Software patterns proposed by Weber et al. [S30] and Liu et al. [S37] open new possibilities for ameliorating smart contract change management. Weber et al. [S30] suggest using *contract factory* defining methods for instantiation of specific case contracts from predefined generic contracts. There are works in progress to handle the smart contract evolution, e.g. *delegation technique* where the data and logic of the smart contract are separated and *registry contract technique* which keeps the track of smart contract versions and records the history of changes and rationales [S13],[S45]. Marino et al. [S15] suggest guidelines for smart contract change management that have been applied for altering and undoing smart contract execution on Ethereum platform. The guidelines that are taken from legal contracts and adjusted to the context of smart contracts, briefly, are (i) modification by right resulting in a new contract, (ii) modification by agreement, and (iii) reformation. Despite this, change management mechanisms for smart contracts have not been fully explored in the context of BBS engineering process. This is an important topic for future research.

—*Dispute resolution design.* The enforcement of smart contracts for business services raises the question about the scope and responsibilities in the case of arising disputations whether they should be resolved automatically by smart contracts or human intervention [S52]. In a business environment, a violation or illegal act against a contract can be resolved by a well-established body of laws such as imposing fines, asset detachment, and service deprivation in courts. On the other hand, despite some debates, violations like malicious security attacks, network disruption, and power cut at software or network levels cannot be fixed without the need for dispute resolution. Such issues imply that software teams should clarify and verify with stakeholders the boundary and scope of user stories and actor interactions that should be converted to smart contracts.

—*Consensus mechanism design*. Validator nodes are responsible for validating a newly added block to the blockchain. The new block is propagated to the network and is appended to the chain of blocks once all validators reach a consensus to ensure that they have an exact copy of the new block. In this regard, the task of consensus mechanism design deals with defining an agreement mechanism satisfying all validators involved in the network when a new block is created. Efficient consensus creation has been an important query in distributed system engineering design and several consensus mechanisms have already been explored over the years. Some are applicable in a BBS context. The choice of prevailing consensus mechanisms such as proof of work (PoW), delegated proof of stake (DPoS), proof of importance (POI), and proof of stake (PoS) [12],[18],[19],[37] requires a software team to make the trade-off between quality factors such as power consumption, scalability, and simplicity. The complexity of the consensus mechanism depends on the network type. A permissioned network needs less complicated mechanisms due to restricted network access and vulnerability to security attacks. On the flip side, in a permission-less network, an energy-intensive mechanism, such as PoW negatively affects BBS scalability and thus reduces the transaction processing throughput in contrast to permissioned networks that tend to utilize non-compute-intensive mechanisms like DPoS.

—*Incentive mechanism design*. Due to the power consumption and computation fee charged in the network to validate new blocks of transactions, some economic rating and reputation mechanisms are defined for validators, e.g., miners. The incentive mechanisms for active validators can be defined for different aspects such as a gas fee for transaction execution, storage price, security deposit, data retrieval, and so on [S54]. Additional economic incentives are needed for validators if BBS has computations that are being run off-chain.

### 5.2.3  BBS implementation and test

The architecture designed in the previous phase is realized by developing software and hardware components. According to the identified papers, a key aspect in conjunction with the implementation and test is the choice of a blockchain platform. While tool support for the whole BBS lifecycle automation may not be feasible, the review of selected studies reveals several popular ones, either open-source or commercial, which are used during this phase for implementing on-chain components. The example of these, as shown in Table 5 and alphabetically sorted, are *BigChainDB*, *Chain Core*, *Corda*, *Credits*, *Domus Tower*, *Elements*, *Ethereum*, *HydraChain*, *Hyperledger Fabric*,

*Hyperledger Iroha*, *Hyperledger Sawtooth Lake*, JUICE, *Multichain*, *Openchain*, *Quorum*, *Stellar*, *Symbiont Assembly*, and *Truffle*. Apparently, the most frequently used platforms are *Hyperledger Fabric* and *Ethereum*. For example, Hyperledger Fabric [65] is an open-source development platform, hosted by the Linux Foundation. It enables to model and integrates existing systems with blockchain platforms. Hyperledger Fabric's programming model, which supports Node.js and Java, enables more intuitive development in a plug-and-play fashion. An analytical comparison of features offered by these platforms is out of the scope of this survey. However, the question has been addressed by Bettín-Díaz et al. [S18] where they present a synopsis of their experience in BBS engineering with an application in supply chain management. The authors recommend that software teams should consider five key factors in choosing a platform, given the need to achieve an alignment between technical and managerial decisions. They are (i) *maturity* of the candidate blockchain platform in terms of how long it has been in the market, supporting model, and availability of documentation and training materials, (ii) *ease of development* which is based on required programming skills to use the platform, (iii) *confirmation time* which depends on the choice of consensus mechanisms, (iv) *security support between nodes* as a platform may provide functionalities to enable public or private network configuration, and (vi) *APIs support* for functions related to authorization, authentication, hash generating, data storage and retrieval, and smart contract lifecycle management. Furthermore, the selected studies suggest that mainstream tools and programming languages are used for the coding, scripting, transferring/streaming, and data manipulation of off-chain components.

Table 5. Sample technologies used during the phase of BBS implementation and test

| | Technology | Type | Aim |
|---|---|---|---|
| On-chain components implementation | BigChainDB | Blockchain platform | A big data distributed platform with a support of blockchain characteristics such as de-centralized control, immutability, and digital data transfer mechanisms |
| | Californium CoAP | Development framework | Securing data transfer between IoT device data and blockchain platforms |
| | Chain Core | Blockchain platform | Issuing and transferring financial assets on a permissioned blockchain infrastructure. |
| | Corda | Open source distributed ledger | Enabling the development of smart contracts with a support for pluggable consensus mechanisms and minimizing transaction cost |
| | Credit | Distributed ledger development framework | Developing permission based smart contracts |
| | Domus Tower | Blockchain platform | Implementing consortium blockchain with a focus on finance domain functionalities |
| | Elements | Open source blockchain platform | Enhancing Bitcoin functionalities at the communication and protocol levels |
| | Ethereum | Blockchain platform | Enhancing Bitcoin functionalities and implementing smart contracts. |
| | Eris:db | Distributed ledger | Enhancing Bitcoin functionalities |
| | HydraChain | Blockchain platform | An extension to Ethereum platform for creating permissioned, private, and consortium blockchain |
| | Hyperledger Fabric | Open source blockchain platform | Providing support for smart contract implementation and test in different application domains |
| | Hyperledger Iroha | Distributed ledger | Developing smart contracts for mobile-based applications |
| | Hyperledger Sawtooth Lake | Open source blockchain platform | Providing support for smart contract development with specific support for decoupling transaction business logic from the consensus layer |
| | JUICE | Tool | Enabling the implementation and monitoring of Solidity smart contracts running on Ethereum platform |
| | Multichain | Open source blockchain platform | Supporting Bitcoin functionalities for multi-asset financial transactions |
| | Openchain | Open source distributed ledger | Issuing and managing digital assets via smart contracts |
| | Quorum | Distributed ledger | Developing smart contract platform based on Ethereum |
| | Stellar | Distributed ledger | Enabling distributed payments infrastructure with RESTful HTTP API servers |
| | Symbiont Assembly | Distributed ledger | A distributed ledger based on Apache Kafka to develop smart contracts |
| | Truffle | Framework | Compilation, test, integration, and deployment of smart contracts |
| | Bitcoin Testnet | Framework | Testing smart contracts without changing real system data or transactions |
| | Hyperledger Besu | Open source development framework | A Java-based Ethereum client to develop and deploy applications to run on the public Ethereum public network or private permissioned network |
| | Mininet | Tool | An emulator to analyse transaction blockchain transaction delays |
| Off-chain components implementation | JSON RPC | Protocol | Remote procedure call used by Ethereum clients to interact with Ethereum nodes |
| | Web3j | Library | A lightweight Java and Android library for working with smart contracts and integrating with Ethereum platform with the minimum overhead for implementing integration codes |
| | CouchDB | Database | A document-based NoSQL database that uses JSON to store data, JavaScript as its query language, and commonly used with Hyperledger Fabric |
| | Raspberry Pi | Toolkit | Collection of hardware and programming language to develop blockchain-based IoT systems |
| | REST API | API | To query data and test HTTP requests as well as call blockchain platform APIs |
| | Bluetooth, ZigBee, WiFi, 2G/3G/4G cellular | Hardware | Hardware Communication protocols |
| | Apache Tomcat, Eclipse Photon and WebStorm | Platform | Hosting and implementing back-end and front-end applications interaction with off-chain components |

Testing is a key engineering task to ensure that an implemented BBS satisfies the specified requirements from the analysis phase. BBS testing that takes into account factors such as type of participants, permissions, input/output states, predefined trigger conditions and response actions of smart contracts, transactions, and expected outcomes in testing scenarios, is very similar to conventional software testing that can be performed at three levels:

—*unit testing* which involves analyzing each BBS component (such as smart contract source code) to identify code segments prone to vulnerabilities during transaction execution and creating the chain of blocks;

—*integration testing* that is to verify if on-chain and off-chain components correctly work together; and

—*user acceptance testing* which is to validate the whole BBS.

Baqa et al. [S44] define a three-step strategy to conduct BBS testing:

—*test map* creates a map of on-chain and off-chain components for which a test is required;

—*test plan* specifies how each type of test, such as unit and integration testing, should be performed within a planned number of test cases. For instance, for the performance test, software teams need to include parameters like measured execution time to record a transaction (time is spent to send a transaction request and receive confirmation from the receiver) and the time taken by cryptographic algorithm execution and consensus creation; and

—*test run* performs test scenarios on a *testnet*, i.e., a testing environment, for example, Bitcoin Testnet, allowing software teams to experiment BBS without worrying about real transactions or using bitcoins.

Conventional software engineering testing techniques such as code coverage review and stress testing may still be applicable in BBS context as acknowledged by [S1],[S44],[S52]. For example, a performance test principally measures the speed of adding a new block to the chain of blocks, and the throughput of the mining, consensus creation, and transaction validation. However, conventional software engineering testing practices appear to be insufficient on their own due to some intrinsic features of BBS as discussed in Section 5.2.2. As mentioned earlier (Section 5.1 and 5.2.2), smart contracts in Ethereum are typically immutable by default. Once they are created and deployed, their source code is infeasible to change as they act as an unbreakable contract among participants. Nevertheless, in some scenarios, it is needed to update smart contracts if an error is detected during the smart contract test or the parties agree to change. Unlike conventional system development, further updates to fix identified bugs in smart contracts can subsequently become a problem. Findings from a survey of 232 smart contract developers, conducted by Zou et al. [S13], highlight the top testing challenges of smart contracts as (i) difficulty of identifying test scenarios, (ii) unpredictable flaws in blockchain platforms, virtual machines, and compilers, (iii) the lack of mature testing frameworks, (iv) asynchronous environments of on-chain and off-chain components, and (v) cost of gas consumption for testing smart contracts. *Code review* is a basic way to ensure the validity of a smart contract. To this end, software teams can employ techniques such as (i) peer smart contract code review, (ii) requesting blockchain practitioners from the GitHub community to check a smart contract, and (iii) asking third-party reviewers to audit smart contract code. Zou et al. [S13], however, identify potential problems with code reviews including speed and the difficulty of finding expert developers to detect security flaws in smart contracts. To lessen the complexity of code reviews, [S51] and [S52] similarly suggest three stages of testing in order to gradually identify and fix bugs in a smart contract code:

(i) deploying smart contracts on a local network, (ii) deploying smart contracts on the test network for software teams to use, and (iii) deploying smart contract on the live and main network and to execute them by users.

Automated testing techniques are essential to reduce the complexity of code review and to identify vulnerabilities in a smart contract before its deployment. In line with this, as mentioned earlier in Section 5.1.2, using MDD enables developers to represent the general features of a system at a high abstraction level without being constrained to platform-specific implementation details. A common technique to verify a smart contract is *model-checking* [66]. Applied in a BBS context, it enables the production of code coverage test scenarios to examine the correctness of function execution, sequence of functions, branches, and trigger points in smart contracts. Model-checking can be accommodated in the implementation phase for the automatic generation of smart contract codes based on system's behavioral state models and non-trivial environmental interactions, e.g., [S31], and consistency checking between the smart contract deployment model and deployment scripts for distributed ledgers, e.g., [S32]. In this spirit, Destefanis et al., [S52] propose a model-based testing technique as a rigorous mechanism for automated smart contract tests where a set of particular fault-related test cases are generated from an abstract fault model. However, the test scalability of automated tests in a large number of smart contracts can be considered as a limitation.

*5.2.4 BBS maintenance*

We found that the discussion on the maintenance phase is scant in the existing literature. This is, perhaps, due to the fact that, the maintenance of BBS has no difference at its core with non-BBS. The only study that provides empirical findings on the maintenance phase is by Bosu et al., [S22], which highlights the challenges that are not noticeable in non-BBS development. In other words, as software teams can upgrade conventional software applications with new features as supported by core technologies like DevOps [52], BBS needs to wait for the preparation of all nodes around the network to upgrade new functionalities which may cause issues such as outdated results, delay in synchronization, and costly response to changes in the business domain. As mentioned earlier, due to the difficulty of making changes in smart contracts after deployment, backward compatibility, the capability of earlier transaction validation, or rolling back to a previous version of smart contracts are challenging.

**5.3 RQ3: What software modeling approaches and notations are applicable in BBS development lifecycle?**

Our perspective on BBS engineering here is divided into *models* and *modeling languages* (Figure 8). Models conceptualize different components of a BBS and *modelling languages* determines a means, i.e., syntax and semantic, to express these models.

*5.3.1 Models (work-products/artefacts)*

BBS engineering can be viewed as series of intermediate models that are generated and evolved during the development process endeavor to achieve a final BBS. The models are results of development process tasks performed by software teams. They help trace how high-level stakeholders' requirements are transformed to executable BBS. The importance of models in BBS context is much comparable to conventional software system development. Models may be cost-prohibitive to generate, either manually or automatically. Due to their evolution and maintenance cost, modeling is situational and it depends on factors such as project requirements, the project domain, and developers' opinions. Hence modeling does not happen in a vacuum.

The literature suggests some notable models associated with the analysis phase. Amongst them, a *requirements model*, which has been frequently stated in the selected studies by different terms such as *user stories* and *goal statement* [S1], *use-case model* [S23],[S31], *collection of requirements* [S31], *system requirements* [S46], and *business model* [S51], collectively, describe functional/non-functional requirements that should be fulfilled by target BBS. It forms a basis for analyzing the remaining domain. Lin et al., [S1] in their experience of developing a security system for the telecommunication domain suggest documenting requirements as user stories where customers write a short description of a certain transaction that is needed to be delivered by BBS in their own terminology and index cards. Marches et al., [S23] propose *use-case models* as a starting point to identify how actors will interact with and send/receive transactions to BBS. *Use-case models* should capture features and functions specific to processing regulations and technical requirements such as legal constraints and cryptographies. Requirements models are compiled in a feasibility report, supplemented by a primitive *prototype*, also referred to as *blockchain compliance checklist* [S51] and *theoretical build-up* [S55], to get a better understanding of requirements and the technical feasibility of BBS. As mentioned earlier, prototypes are not only meant for requirements elicitation but also provide a foundation for BBS architecture models.

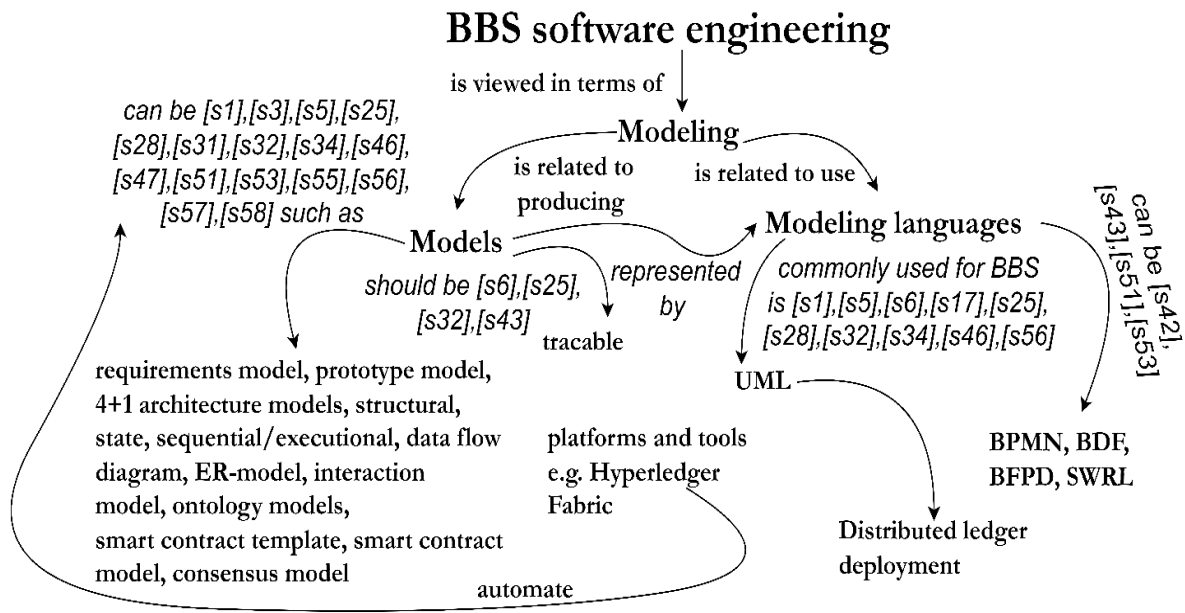

Fig. 8. *Modeling* conceptual map

The literature recommends some models related to the design phase of BBS engineering. Recall from 5.1.3, the set of well-known software engineering *architecture models*, 4+1 views [58], are used in the context of BBS design phase. They describe the structural relationship and interaction between software and hardware components of BBS. For instance, a *process view* of the architecture shows the sequence of events occurring when users or off-chain components interact with on-chain components such as a smart contract function call (Figure 5). Similarly, the *deployment model*, one of the commonly recommended 4+1 models, as noted by Górski and Bednarski [S32], represents how (i) BBS off-chain components access on-chain components and (ii) BBS architecture is operationalized with respect to a chosen blockchain platform, physical layers, and installation scripts to make BBS available to end-users. Additionally, Rocha and Ducasse [S53] suggest the usefulness of entity-relationship models (ER-model) to represent conceptual and logical design for blockchain data exchange between on-chain and off-chain components.

Apart from the well-known models that are recommended above, software teams may generate some blockchain-specific models during the development process. A smart contract model, as recommended in [S1],[S5],[S31],[S46],[S56], [S57], represents the functionalities that realize requirements that are indicated in BBS analysis phase.

The rationale for the applicability of object-oriented modeling in BBS is that a smart contract model can be analogically viewed as a *Class*. Due to different options to call their function, smart contract models are classified into *structural*, *state*, and *sequential/executional*. These three different models are to represent the structure and relationships, transition states, and sending/receiving messages from/to a smart contract, respectively. To be more specific, a structural model, akin to a class model, shows data, public, and private functions of a smart contract, and allows for reuse and inheritance from other smart contracts. A state model enables developers to identify the different situations of a smart contract in its whole life cycle from the stage of deployment to sign-in, execution, and termination. As such, different changes in the status of a smart contract, e.g., chaining of transactions, as results of internal operations or interactions with off-chain components such as external systems can be identified. The other useful model at the design phase is a sequence model. It represents interactions in the form of messages from external systems/actors to BBS or from smart contracts to other smart contracts. The information obtained on interactions between BBS components via a sequence model may help refine architecture model and smart contract model [S1], [S3],[S23]. Bettin-Diaz et al [S18] suggest *data flow diagram* to identify how data is passed across running off-chain and on-chain and where it is processed.

As the deployed smart contracts cannot call external APIs, interfaces are implemented via front-end web technologies, such as JavaScript and HTML5, to enable on-chain components to interact with off-chain components. In this regard, an *interaction model* is useful for identifying important behavioral patterns of interactions and thus defining the integration points that should be implemented for connecting on-chain and off-chain components [S28],[S34],[S47],[S56].

A blockchain-specific work-product during the design phase, as a result of *consensus mechanism design* task (as defined in section 5.2.2), is the *consensus model*. This model shows how an agreement between all nodes in BBS is reached to accept a new block to the blockchain. A consensus model reflects design decision parameters such as transaction throughput, latency, network bandwidth, rules, and incentives that should be enforced by a selected consensus mechanism.

Since software teams generate models for a purpose, the traceability and dependency among these models, i.e., what models are predecessors/successors or master-subordinate to a certain model, in the whole BBS development process endeavor is intuitive. This is in line with the discussion of software requirements management and traceability. Gotel and Finkelstein [S58] explain that traceability is the ability to follow requirements in both forward and backward directions starting from their origins towards implementation, deployment, and maintenance phases. According to this line of argument, establishing the traceability among models in a BBS development endeavor enables better development process management, automatic forward and backward engineering, responsibility assignment to teams, and error detection. The review of the selected set of studies shows that traceability is weakly supported in the literature. Only a few works define the chain of models for the development process: [S6] (*smart contract template → smart contract model→ code model*), [S23] (*use-cases model* [main actors of BBS] → *class model* [BBS relationships with actors] → *sequence model* [realization of transactions by BBS], [S32] (*smart contract model → code model*, *deployment model→ configuration model*), and [S43] (*ontology models → smart contract model*).

### 5.3.2 Modeling languages

The notations and semantic rules imposed by a modelling language that is used during a BBS development endeavor offer threefold advantages (i) enabling precise expression of BBS aspects and generated work-products, (ii) providing a consistent way of communication between software teams and stakeholders, and (iii), increasing the automation level of code generation and test.

From the reviewed studies, it is found that Unified Modelling Language (UML) [67]– adopted by the Object Management Group (OMG) in 1997 as a de-facto for object-oriented modelling– is the most often adopted modeling language for BBS development [S1],[S5],[S6],[S17],[S23],[S28],[S32],[S34],[S46],[S56]. An advantage of using UML class diagrams to represent the internal structure of a smart contract such as functions and data attributes is automatic smart contract code generation. Extending UML through stereotypes to cater necessary BBS-specific modeling requirements in BBS development is intuitive [S46]. The lack of capability for modeling different types of distributed ledger technologies has motivated Gorski and Bednarski [S32] to propose UML stereotypes and tagged values for generating deployment scripts and configuration files, organized as UML profile for *distributed ledger deployment*. Apart from UML application, it can be seen that BPMN (Business Process Modeling Notation) [68], a de-facto for flow-oriented representation of core constructs of business processes, is used as a complementary to UML for business people. In this regard, a first promising attempt by Rocha and Ducasse [S53] suggests that using BPMN in BBS development process has advantages to identify (i) transactions, functions, and actors in collaborative business processes to map them to smart contracts and (ii) interactions –integration paint points– between off-chain and on-chain components. For example, the *swimlane* notation, which is the named box container, can be used to specify interactions between legacy systems and BBS. Other alternative notations to BPMN for the same purpose of business processes modeling, suggested by Almeida et al. [S51], are namely BDF (Block Diagram Flow) and BFPD (Block Flow Process Diagram).

Recall from Section 5.1, it tends to be seen that ontologies and related technologies including Web Ontology Language (WOL) and Semantic Web Rule Language (SWRL) are suitable for modeling smart contracts [S42],[S43]. In their work, Choudhury et al., [S43] provide a modeling framework for the automatic generation of smart contracts. The framework adopts ontologies to represent and encode business constraints and semantic rules available in unstructured business documents (legal contracts, tables, and charts). This standard representation via SWRL rules ensures accurate parsing and enables the automatic generation and test of smart contract templates in a given BBS application domain.

### 5.4 RQ4: What are the key roles in a BBS development endeavor and how do they play?

Theoretically, a BBS development endeavor can be centered on the view of bringing people who collaborate to reach the final BBS product. They can be either producers in a software team who are responsible for creating, assessing, iterating, maintaining models/work-products or end-users who interact with BBS. BBS engineering relies on the availability of roles with technical expertise, business acuity, and well-defined skill portfolios that should be acquired and ensured that they have a clear understanding of their responsibilities. Development roles need to be tailored in accordance with the project settings. For example, the development of a BBS for public blockchain, as pointed out in Section 5.2.2, requires various BBS-specific roles associated with designing, testing, and deploying smart contracts to potentially unknown users participating in a peer-to-peer distributed blockchain network. Unlike a limited number of roles that are involved in a BBS development for a private blockchain type, the more cooperative effort might be required for a BBS that is developed for a public blockchain type. For example, broad arrangements between parties about the content and objectives of smart contracts should be made to ensure that preconditions are met before smart contracts are turned into executable codes.

Apart from the commonly-known roles in non-chain software engineering such as project manager/technical leader developer, network administrator, and data modeler, as equally emphasized in BBS [S1],[S13], there are a few new roles introduced in BBS engineering. As shown in Figure 9, they are broadly split into five distinct groups of (i) *core blockchain developer*, (ii) *blockchain software developer*, (iii) *systems integration engineer*, (iv) *legal professional*, and (v) *blockchain user* [S1],[S6],[S13],[S45],[S52],[S55],[S56]. All these roles collaborate to accomplish the development of BBS, yet each role is associated with different responsibilities during the development lifecycle.

The role of a *core blockchain developer* refers to a development party who is responsible for designing blockchain platforms, APIs, protocols, network architecture, and security patterns related to blockchain technology. The *blockchain software developer* role, on the other hand, utilizes enabling foundations provided by the core blockchain developer to implement BBS running on blockchain platforms. Typically, a blockchain software developer is responsible for writing smart contracts that codify critical business logic in a secure way, implementing interactive front-end interfaces with BBS, backend systems, and maintaining the full stack of running BBS. Bosu et al. [S22] highlight three essential skills for blockchain software and core developers: (i) strong security programming background, (ii) networking knowledge for secure design of communication between distributed off-chain and on-chain components, (iii) knowledge of mathematical cryptography for designing BBS algorithms and protocols. In conjunction with these skills, in-depth knowledge about scalable architecture design is an essential skill for the role of developers [S1],[S5]. *Systems integration engineers* play an important role. She/he is in charge of integrating all on-chain and off-chain components, assuring the satisfaction of interoperability requirements, seeking approvals of component changes, and ensuring the entire BBS functions correctly [S16].

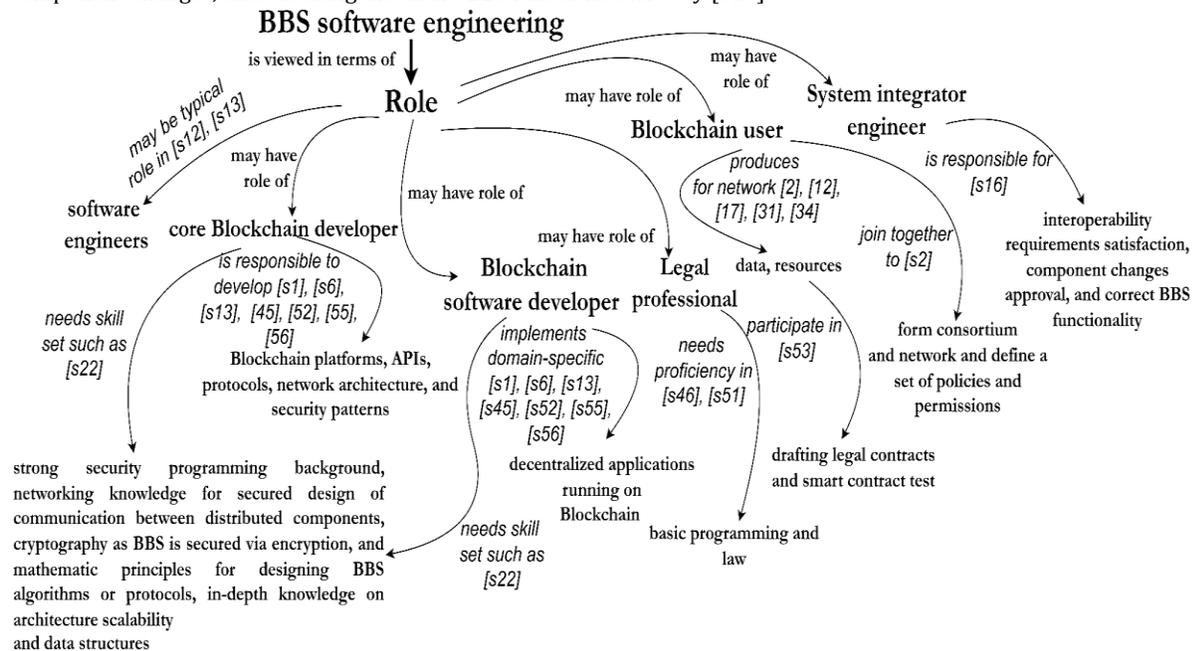

Fig. 9. *Role* conceptual map

Apart from the abovementioned technical roles, BBS engineering may entail a portfolio of roles with expertise in finance and law domains. *Legal professionals* are needed to interface between business-focused contractors and technical-focused software teams. One of responsibilities in this role is to identify potential contractors and partners. Giancaspro et al. [S46] and Almeida et al. [S51] highlight a need for a new generation of IT lawyer roles in software teams for complex business transactions deployed on blockchain. The role that comes into play across different phases of the development process should have basic proficiency in programming and be responsible for drafting legal contracts and ensuring their clauses are adequately translated into relevant smart contract code. Rocha et al. [S53] point out that legal expertise is crucial to ensure test case coverage and to verify the prose of a legal contract in smart contracts.

Finally, the role of *blockchain user* in a business model integrated with a blockchain eco-system is to consume or produce data in blockchain and contribute to resources that are beneficial to others [S2],[S12],[S17],[S31],[S34]. Blockchain users, join to form a network and define a set of policies and permissions agreed to by all in a consortium or private blockchain type architecture.

## 6 FUTURE RESEARCH DIRECTIONS

The review of the literature in Section 5 not only presents an overview of concepts and the fundamental understanding of BBS engineering that is portrayed by our conceptual framework, but also reveals the symptomatic needs for holistic and integrated engineering approaches. The situation, literally, requires further research to ameliorate the status quo of software engineering for BBS. The following research gaps, which are related to each aspect of our conceptual framework, deserve further exploration.

### 6.1 Absence of development process tailoring

It has been well-recognized that software engineering approaches should be tailored to circumstances surrounding a project. The essence of this is truly stated by Sommerville and Ransom [69], p.93: *it is a truism that any [development] method/approach has to be adapted for the particular circumstances of use*. The view that Brook [70] calls the *silver bullet (or one-size-fits-all assumption) is not a practical choice* continues as a persistent theme in BBS engineering. In recognition of this, Miraz and Ali [71] examine the suitability of adopting conventional software development practices in a BBS context. They conclude that software teams may omit, for example, certain tasks in a BBS development endeavor not from a position of ignorance, but from the pragmatic reason that those tasks are not relevant to the project's characteristics, functions of a selected blockchain platform, or the choice of blockchain type. No matter how well-crafted, it might be impractical to find or design a single top-down engineering approach applicable to all BBS projects and in totality as *imposed from above*. Software teams may have their in-house approach for software development; however, that approach may confine its focus on some BBS development tasks and ignore others. For example, in the early phases, if a software team seeks recommendations on the feasibility of a BBS instance in a certain organization, the findings in [S24],[S46],[S47] are quite relevant to be considered by the team. On the other hand, if the team is rather in the design phase and wants to know about smart contract design techniques, then a piece of advice in the work by Frantz and Nowostawski [57] for high-level design as well as work by [S31], [S46],[S47] for low-level design is applicable.

Given the specific situational factors and architectural decisions of BBS development context, as some listed in sections 5.2.1 and 5.2.2, one may logically conclude that BBS engineering approaches should be malleable or selected from a portfolio of approaches. A suggested solution in the software engineering discipline, which can be equally applied in BBS engineering, is *situational method engineering (SME)* [72]. SME tailors/creates project-specific engineering methods/approaches via selecting appropriate method fragments from a method base and assembling them to construct a highly customized method for a given project. However, to the best of our knowledge, little research exists on process tailoring for BBS development. Questions like *what key method fragments, e.g., process patterns* (Section 5.1)*, are required to be populated in a method base?* and *how can they be assembled according to the project situation factors to tailor or create a custom-specific methodology?* are yet to be addressed in more depth.

### 6.2 Requirements analysis is a dire problem

Whilst the topic of requirements analysis has achieved its maturity in conventional software engineering, it is currently not studied as much in BBS engineering. Apart from blockchain-specific characteristics, the requirements analysis for BBS does not significantly differ from any non-BBS. Recall from Section 5.2.1, software teams can still use the conventional requirements analysis techniques, as suggested by [S5],[S31],[S49]. On the contrary, the selected studies overlook questions regarding what requirements contribute *value* to BBS stakeholders. In other words, this is left as an unexplored area in BBS requirements analysis that what requirements are valuable and they instigate a variety of stakeholders with diverging goals and commitment levels if they are addressed by BBS, for example, smart contracts. This argument is also confirmed by the case study findings in Du et al., [S47] who accentuate the complexity of BBS for stakeholders and the fact that they primary care if BBS is useful rather than what the technology hype offers. As the definition of smart contracts and solution architecture depend on the elicitation of true requirements as well as legal contracts, a tighter connection between the analysis phase and the design phases is required to ensure smart contract errors are detected and fixed at the early stage of BBS development. More empirical studies are needed to get a better understanding of the criteria and techniques for selecting requirements whose fulfillment by BBS will create a truly added value for stakeholders.

### 6.3 Importance of model traceability

Some selected core papers (cf. Section 5.3) state that the creation of models has a vital role in BBS development to manage complexity, represent information at a different level of abstraction and increase the automated model generation when the scale of transactions is large. However, few studies such as [S6],[S25],[S32],[S43] establish a chain of models to be produced to reach final BBS. Leveraging well-established state of the art in MDE based software engineering [54], as pointed in Section 5.3.1, keeping predecessor-successor or master-subordinate traceability and relationships between models in BBS engineering is important to enable (i) the evaluation of the extent to which legal contracts and security requirements are matched with executable smart contracts, (ii) the assessment of the degree to which each component in a BBS development justifies its existence, for example, whether a smart contract's function is traced to the prose of a legal contract that it satisfies, and (iii) automatic forward and backward model transformation of smart contracts. Embarking MDD based engineering approaches can be helpful in a seamless and automated transformation of intermediate BBS models. This is an important avenue for future research.

### 6.4 Legacy system blockchain enablement is weakly supported

Legacy software systems operating and storing critical organizational data may predate blockchain technology. The term legacy system is often associated with older technologies such as mainframes, monolithic architecture, file systems, communication protocols, and programming languages such as FORTRAN, COBOL, and C. A software system might have been developed using more recent technologies such as Microsoft .Net and J2EE. However, it may not support blockchain specific design considerations such as immutability, stateless/stateful smart contracts, or integration between off-chain and on-chain components (Section 5.2.2). Such a system is considered a legacy if it is going to be migrated to a blockchain platform. Software re-engineering of these systems to make them blockchain-enabled is a crucial endeavor. This is due to reasons such as required data migration, interoperability between blockchain and legacy platforms, refactoring of legacy system code to smart contracts, mode of hosting, blockchain platform vendor lock-in, and security issues of transparency and openness to participant nodes in a blockchain network. These imply that the migration of large scale and complex legacy systems to blockchain platforms needs to be organized and anticipated in a more systematic way.

Thus, it is worth investigating how and what new practices a software team should incorporate into the development process to make legacy systems blockchain-enabled. Research exists on legacy system migration to SOA [30],[73] and cloud computing platforms [27],[31],[74]. It is time to help continue this research stream to the topic of migrating legacy systems to blockchain platforms. The presented framework in this survey, in light of the reported findings in [S38], is a starting point to set the scene for further explanation.

### 6.5 Need for stakeholder-driven engineering approaches

Presumably, the support of all stakeholders, amongst others, is a key aspect of any successful software engineering project. This is equally important in the context of BBS engineering projects. Recall from Section 5.4, BBS engineering introduces new types of stakeholders as opposed to non-BBS engineering. These range from core blockchain developers to legal IT professionals, who may have multiple roles within the development lifecycle and have an interest in (or are influenced by) the lifecycle. In light of our general guidelines for BBS stakeholders presented in Section 5.4, a stakeholder-driven perspective for BBS engineering with an emphasis on roles and distributed teams/development is beneficial to improve the project management and governance through the entire development lifecycle. Researchers can also run empirical studies across multiple teams and organizations to identify *when* each stakeholder is involved, *what* roles are assigned, *where* and *how* stakeholders influence and are influenced in the course of analysis, design, implementation, and operation of BBS. The results of these studies can contribute to the identification of stakeholder-driven BBS engineering approaches that are significant in the areas of (i) BBS requirements analysis and design phase, (ii) BBS adoption in organizations, and (iii) project performance measurements.

## 7 THREATS TO VALIDITY

We divide the limitations of our survey in terms of *internal* and *external* validity. The former bears on factors that we might have been unaware of or unable to control them whilst the latter concerns with threats that negatively affect the generalizability of our survey findings.

### 7.1 Internal validity

As far as the internal validity is concerned, due to a sheer volume of published studies in academic and multi-vocal literature on blockchain, it was infeasible to conduct an exhaustive literature search to identify all papers in the digital libraries related to RQ-RQ4 stated in Section 4.1. One reason is the inclusion criterion in conducting our SLR as described in Section 4.1. We have given a priority to identify well-cited papers with validation on real-world applications from the literature in order to increase the reliability of our presented conceptual framework and reported survey findings. We have been reluctant to consider opinion papers, technical reports, white papers, and books in our survey. Moreover, we observed that some important papers that had not directly discussed the software engineering of BBS; nevertheless, their identified challenges and propositions were truly related to the focus and scope of this survey.

Besides, the blockchain field is still an immature area and its literature is overwhelmed with interpretations and variants of concepts that might not necessarily be homogeneous. We found that some papers do not use the term of software engineering in any section, but they were important in crafting our framework and sharing important challenges and proposed countermeasures related to the subject of this survey. For instance, any of our search queries combining the key terms like *software engineering* and *blockchain* over the digital libraries directly resulted in identifying papers such as [S26], [S30], as they are referred to in Section 5.1 and classified under the design phase (Section 5.2.2). Inversely, their inclusion in the selected papers was a result of several rounds of the snowballing technique. Unavoidably, the selection of the papers might have been influenced by subjective bias. To lessen this issue, two co-authors conducted the literature review, paper selection, content analysis, and framework generation whilst senior researchers oversaw the whole survey.

We conducted the purposeful snowballing technique and the careful scanning of the identified studies from the literature to select the ones that strictly satisfied the selection criteria (Section 4.1). An observation in the selected studies reveals that the highest number of the studies are affiliated with Australia as same as the authors' affiliation. The reason is that we started with an initial set of papers from our domestic academic colleagues and expanded them to identify new papers through reviewing the related studies and references that were cited in this set. This may have led to a potential bias in our resulting papers set as listed in [17].

### 7.2 External validity

As for the external validity, we do not claim to provide complete coverage and inclusivity of the presented conceptual framework. A less discussed aspect in Section 5.2.3, yet important, in the presented conceptual framework, is the *tool* support for BBS. It is associated with the aspects of *process* and *modeling* in the framework. As our survey intends to provide an overarching layer for BBS development by abstracting pure technical and dispersed literature, we believed that adding the new aspect of *tool*, despite the availability of some dedicated BBS tools (Section 5.2.3), can be either a discussion on the way of operationalization of the framework's fragments or a topic of future surveys based on our proposed framework. Finally, whilst the derivation of the framework through Grounded Theory has been iterative and with several refinements in the literature source and coding procedures, we don't claim about the generalizability of the presented framework and accounted findings beyond 58 source papers in the current survey.

## 8 CONCLUSION

We presented an SLR on software engineering for BBS. We addressed the most relevant and up-to-date survey research questions synthesizing published material available in the blockchain literature. We explored the current state-of-the literature and identified important future research directions in light of the fundamental aspects of the conceptual software engineering framework for BBS. In terms of the process aspect, firstly, the discussion on requirements analysis has received much less attention from researchers. Further research can investigate how or which added-value requirements should be selected to realize by BBS. Secondly, some researchers have proposed situational characteristics upon which BBS development process should be tailored and enacted. Whereas the need for a flexible and bespoke BBS development process is acknowledged, there is a dearth of research dealing with this issue and little is known about how tailoring is applied for BBS engineering projects. Thirdly, considering the importance of legacy software systems supporting back-end functions of IT-based organizations and storing critical data, the minority of the identified papers discuss legacy software re-engineering to blockchain, for example, steps to identify and integrate core business logic, e.g., Web-Service, to smart contracts. Migrating legacy software to blockchain has recently received attention. A future trend may particularly focus on software re-engineering for making legacy systems blockchain-enabled. Fourthly, in terms of the modeling aspect, several models are generated during a BBS development lifecycle ranging from requirements to smart contract design elements, and down to smart contract code segments. Managing the chain of these models as a means to understand logical traceability and bounded dependency among these models plays an important role to enable automatic and tool-based BBS design in view of stakeholders'' requirements. Techniques related to the automatic traceability of models in BBS development are partially captured in the literature, which stimulates another possible future research. Fifthly, from the aspect of stakeholders, bearing in mind that BBS development involves new types of technical and non-technical stakeholders than many conventional software systems, with diverse goals in using BBS and different levels of engagement, stewardship, cooperation, and incentives, this survey provides opens a new research stream on stakeholder-driven BBS development.

The eventual upshot is that software engineering for BBS is not as straightforward as for many non-BBS ones. Contaminated with partiality and subjectivity, adopting an ad-hoc approach is hardly safe and secure for business-critical BBS. More research is necessary to move from ad-hoc BBS development to more disciplinary, repeatable, and controllable BBS engineering and to enhance the maturity of this research field. To pave this path, we explored the current research state of the literature and identified important future research directions in light of a new conceptual framework. We hope that researchers benefit from highlighted ideas that are shared in this survey as entry points to tackle unaddressed challenges. The practitioners can incorporate the findings in this survey into their in-house software development approach to enhance its capability to support BBS development endeavors.

**Acknowledgement**. The authors would like to thank feedback from associate editor and reviewers. The work of Professor John Grundy is supported by ARC Laureate Fellowship FL190100035.